\documentclass[aps,prb, superscriptaddress,amsmath]{revtex4}

\usepackage{graphicx}
\usepackage{psfrag}

\newcounter{saveeqn}

\makeatletter

\def\@dotsep{4.5}

\makeatother

\renewcommand{\arraystretch}{0.5}

\begin{document}

\title{\Large{Eliminating fast reactions in stochastic simulations of biochemical networks: a bistable genetic switch}}
\author{\large{Marco J. Morelli}}
\affiliation{FOM Institute for Atomic and Molecular Physics, Kruislaan 407, 1098 SJ Amsterdam, The Netherlands}

\author{\large{Rosalind J. Allen}}
\affiliation{SUPA, School of Physics, The University of Edinburgh, James Clerk Maxwell Building, The King's Buildings, Mayfield Road, Edinburgh EH9 3JZ, UK}

\author{\large{Sorin T\u{a}nase-Nicola}}
\affiliation{FOM Institute for Atomic and Molecular Physics, Kruislaan 407, 1098 SJ Amsterdam, The Netherlands}


\author{\large{Pieter Rein ten Wolde}}
\affiliation{FOM Institute for Atomic and Molecular Physics, Kruislaan 407, 1098 SJ Amsterdam, The Netherlands}

\date{\today}


\begin{abstract}

\linespread{1.5}
\large{

  In many stochastic simulations of biochemical reaction networks, it
  is desirable to ``coarse-grain'' the reaction set, removing fast reactions while
  retaining the correct system dynamics. Various coarse-graining
  methods have been proposed, but it remains unclear which methods are reliable and which reactions can
  safely be eliminated. We address these issues for a
  model gene regulatory network that is particularly sensitive to dynamical 
  fluctuations: a bistable genetic switch. We remove protein-DNA and/or protein-protein
  association-dissociation reactions from the reaction set, using various coarse-graining strategies. We determine the effects on the steady-state probability distribution function
  and on the rate of fluctuation-driven switch flipping transitions.  We find that protein-protein interactions may be safely eliminated from the reaction set, but protein-DNA interactions may not. We also find that it is important to use the chemical
  master equation rather than  macroscopic rate equations to compute effective propensity functions for the coarse-grained reactions.

}

\linespread{2}
\end{abstract}

\maketitle
Current address for Sorin T\u{a}nase-Nicola: University of Michigan, Physics Dept.,  450 Church St., Ann Arbor, MI 48109-1040, USA\\
\newpage

\section{\large{Introduction}}\label{sec:intro_cg}
\large

Biochemical reaction networks control how living cells function. Computer simulations
provide a valuable tool for understanding how complex biochemical network
architecture is connected to cellular function. A popular method for simulating biochemical networks is the ``Stochastic
Simulation Algorithm'' (SSA) which was introduced in this field by
Gillespie \cite{Gillespie76,Gillespie77}.  For many reaction networks, however, SSA simulations are prohibitively expensive because of
``time-scale separation'': the reaction set contains some reactions which occur much more frequently than others. For every ``slow'' reaction
event, many ``fast'' reaction events have to be simulated.  This problem has led to the development of
various methods for coarse-graining the reaction set
\cite{Haseltine02,Shibata03,Bundschuh03,Rao03,Puchalka04,Cao05,E05,Salis05} - that is,
eliminating the fast reactions and simulating only the slow reactions. Key issues are which fast reactions can safely be
eliminated, and how this
should be done, so as not to disturb the original dynamics. In this paper, we address these issues for a biochemical
network which is especially sensitive to dynamical fluctuations: a bistable genetic
switch. Because of its sensitivity, this model provides
a useful test system for assessing how to coarse-grain biochemical
networks.  We expect
our conclusions to be valid for a wide range of biochemical
networks where fluctuation-driven processes are important.

The SSA is a kinetic Monte Carlo method which generates trajectories for the numbers of molecules of each chemical species in the reacting system. The molecular discreteness of the reacting species is included, and the method includes stochastic
fluctuations in the numbers of molecules, assuming that each
reaction is a Poisson process. It is assumed that the system is well-stirred - {\em{i.e.}} the spatial distribution of the components is
ignored (alternative methods that do include spatial effects have recently been developed \cite{Elf04, Ander04, VanZon05,VanZon05_2}). The
SSA generates trajectories that are consistent with the chemical master equation and it has been widely applied in the field of biochemical
networks. A detailed explanation of the method can be found in Refs. \cite{Gillespie76,Gillespie77}. We note that 
some authors include a factor 2 in the definition of propensities for second order reactions. We choose not to use this factor 2.

The SSA becomes inefficient when some of the reaction channels (``the
fast reactions'') have much higher propensities than others (``slow
reactions'') - this is known as the time-scale separation problem, and
several methods for dealing with it have been proposed. In all cases,
the first step is to identify which reactions are ``fast'' and which
are ``slow''. The criterion is generally that fast reactions should
reach a steady state faster than the waiting time between slow
reaction events. The slow reactions are generally simulated using the
SSA, while the various methods differ in their treatment of the fast
reactions. In one class of methods, the fast reactions are propagated
using the deterministic or chemical Langevin equation
\cite{Haseltine02,Kiehl04,Takahashi04,Shibata03}, or via the $\tau$ leap method
\cite{Gillespie01,Puchalka04}; these schemes require that the species
in the fast reactions are present in large copy numbers. In another
class of methods, which we consider here, the temporal evolution of
the fast reactions is given by the chemical master equation.  Because
these algorithms take account of molecular discreteness, they do not
require the species in the fast reactions to be present in large copy
numbers \cite{E05,Salis05,Bundschuh03,Rao03,Shibata03,Cao05,Salis05_2,Kaznessis05}.  These schemes are
discussed in more detail in section \ref{sec:background}.

In this paper, we compare various coarse-graining strategies using a
bistable genetic switch. Bistable genetic switches are excellent model
systems for testing the accuracy of coarse-graining schemes. The
reason is that the spontaneous flipping from one stable state to the
other is driven by fluctuations---bistable switches are thus
particularly sensitive to random fluctuations and hence to
coarse-graining, which tends to remove fluctuations from the
system. Moreover, these switches exhibit dynamics over a wide range of time scales, making them ideal for testing the assumption of time
scale separation that underlies all coarse-graining schemes. The dimerization of
proteins and the binding of dimers to the DNA occurs
on relatively fast time scales of seconds to minutes, while the
synthesis and degradation of proteins occurs on relatively slow time
scales of tens of minutes or longer. Another important time scale
is given by the duration of the switching trajectories, which involve
a combination of DNA binding, dimerisation and protein production
and decay events. Genetic switches
also show dynamics on an even longer time scale, which is the
waiting time between switch flipping events. Since dimerisation and
DNA binding occur on much faster time scales than the inverse
switching rate, one might expect that these reactions could be
integrated out without affecting the switching dynamics. On the other
hand, previous work has shown that DNA binding fluctuations can
actually drive the flipping of genetic switches
\cite{Allen05,Walczak05}. It is thus unclear whether
these reactions can safely be eliminated in simulations of genetic
switches.

We investigate the consequences of integrating out dimerisation and
DNA binding for the switching dynamics of a genetic toggle switch
that consists of two genes that mutually repress each other. A simple
mathematical model of such a toggle switch has been studied by several groups
\cite{CA,Kepler01,Warren04,Warren05,Allen05,Ushikubo06,Lipshtat06}. Using
the SSA in combination with the recently developed ``Forward Flux
Sampling'' (FFS) method for rare event simulations
\cite{Allen05,Allen06_1,Allen06_2}, we compute the steady-state distributions of
all species and the rate at which the network undergoes spontaneous
fluctuation-driven flips between its two stable states. We use these
quantities as measures for the accuracy of various coarse-graining
strategies for eliminating both DNA binding and dimerisation; the
strategies differ in whether the deterministic macroscopic rate
equations or the stochastic chemical master equation is used to
compute averages over the fast reactions. 

Our results show that it is essential to use the chemical master equation
approach rather than the macroscopic rate equation. This is perhaps not
surprising, since the DNA and proteins are present in low copy
numbers. In addition, we find that the steady-state
distributions are quite robust to integrating out dimerisation and
protein-DNA binding; in particular, the location and the shape of the
peaks of the bimodal steady-state distributions are hardly affected by
eliminating these fast reactions from the reaction set. In contrast,
coarse-graining the dimerisation and DNA binding reactions can
have a dramatic effect on the flipping rate. These results can be
understood by noting that the steady-state distribution is
predominantly due to the dynamics {\em within} the basins of
attraction corresponding to the stable states. Within these basins,
the time scales of DNA binding and dimerisation are typically much
faster than those of protein production and degradation; the
time-scale separation requirement is thus satisfied most of the
time. In contrast, the switching {\em rate} is determined by how the
system {\em leaves} a basin of attraction. This relies on rare fluctuations of the fast reactions: in order for the switch to flip, the minority
species has to dimerize and subsequently bind to the DNA target site. Indeed,
as we discuss in more detail in a forthcoming paper \cite{morelli_prep}, during the flipping trajectories, the dynamics of DNA binding and
dimerisation typically do not relax to a steady state.
Thus, the reliability of a particular coarse-graining scheme strongly depends on the quantities one is interested in:
equilibrium properties will typically show a low sensitivity to approximation procedures, 
whereas for fluctuation-driven properties, small errors will be amplified, leading to incorrect results.

In the next section, we describe the model genetic switch. In section \ref{sec:background},  we give background information on the various coarse-graining schemes, and in section \ref{sec:coarse-grain}, we discuss these in the context of the model switch. In Section \ref{sec:Results}, we present results on the accuracy   of the various coarse-graining procedures, using
the stationary distribution and the switching rates as read-outs, and their computational speed-up. We
end with a discussion on the implications of our findings for the
simulations of complex biochemical networks.

\section{\large{The Model Genetic Switch}}
\label{sec:GS_cg}
\begin{figure}
\begin{center}
\includegraphics[width=0.5\columnwidth]{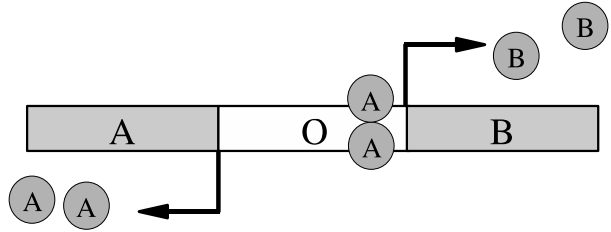}
\caption{Pictorial representation of our model switch, corresponding to Table \ref{tab:reactions}. Two divergently-transcribed genes $A$ and $B$
are under the control of a single regulatory binding site $O$. Proteins A and B both form homodimers, each of which can bind to $O$ and block transcription of the other gene.}
\label{fig:diagram_cg}
\end{center}
\end{figure}

The model bistable genetic switch is shown schematically in
Figure \ref{fig:diagram_cg} and the set of reactions  is listed in Table \ref{tab:reactions} \cite{CA,Warren04,Warren05}.


\begin{table*}

\begin{tabular}{cccccc}
&$\quad\qquad \mathrm{Reaction} \quad\qquad$ & $\qquad\quad\mathrm{Propensity}\qquad\quad$ & $\qquad\quad\mathrm{Reaction}\quad\qquad$  & $\quad\qquad\mathrm{Propensity}\quad\qquad$ & \\
\hline
{\rm Dimerisation} &$\mathrm{A}+\mathrm{A} \rightleftharpoons \mathrm{A}_2$  & $k_{\rm f}n_{\mathrm{A}}(n_{\mathrm{A}}\!-\!1),\quad k_{\rm b}n_{\mathrm{A}_2}$
&$\mathrm{B}+\mathrm{B} \rightleftharpoons \mathrm{B}_2$  & $k_{\rm f}n_{\mathrm{B}}(n_{\mathrm{B}}\!-\!1),\quad k_{\rm b}n_{\mathrm{B}_2}$ &a\\
{\rm DNA\:binding} &${O} + \mathrm{A}_2 \rightleftharpoons {O}\mathrm{A}_2$  & $k_{\rm on} n_{{O}}n_{\mathrm{A}_2},\qquad k_{\rm
off}n_{O\mathrm{A}_2}$  & ${O} + \mathrm{B}_2 \rightleftharpoons O\mathrm{B}_2$  & $k_{\rm on}
n_{{O}}n_{\mathrm{B}_2},\qquad k_{\rm off}n_{O\mathrm{B}_2}$ &b\\
{\rm Production} &${O} \to {O} + \mathrm{A}$ & $k_{\rm prod}n_{{O}}$ & ${O} \to {O} + \mathrm{B}$ & $k_{\rm prod}n_{{O}}$ &c\\
{\rm Production} &${O}\mathrm{A}_2 \to {O}\mathrm{A}_2 + \mathrm{A}$  & $k_{\rm prod}n_{O\mathrm{A}_2}$  &
${O}\mathrm{B}_2 \to {O}\mathrm{B}_2 + \mathrm{B}$ & $k_{\rm prod}n_{O\mathrm{B}_2}$  &d\\
{\rm Degradation} &$\mathrm{A} \to \emptyset$ & $\mu n_{\mathrm{A}}$  & $\mathrm{B} \to \emptyset$  & $\mu n_{\mathrm{B}}$ &e\\
\end{tabular}
\caption{Reactions and propensity functions for the model genetic switch.\label{tab:reactions}}
\end{table*}



As shown in Figure \ref{fig:diagram_cg}, two genes $A$ and $B$ are transcribed in divergent directions, under the control of a single DNA operator region, $O$, which contains one binding site. Coding region $A$ encodes protein A, while coding region $B$ encodes protein B. Both proteins A and B are transcription factors, which, upon homodimerisation, are able to bind to the operator sequence $O$. When $\mathrm{A}_2$ is bound at $O$, the transcription of $B$ is blocked [$\mathrm{A}_2$ is a repressor for $B$]; while, conversely, when $\mathrm{B}_2$ is bound at $O$, the transcription of $A$ is blocked [$\mathrm{B}_2$ is a repressor for $A$]. When neither $\mathrm{A}_2$ or $\mathrm{B}_2$ is bound at $O$, both $A$ and $B$ are transcribed at the same average rate $k_{\rm prod}$. Protein monomers are also removed from the system with rate $\mu$, modelling active degradation processes as well as dilution due to cell growth. For convenience in this model system, we use the rather high value $\mu=0.3k_{\rm prod}$, corresponding to the case where removal from the cell is dominated by active degradation. In our model, we assume that all the steps leading to production of a protein molecule (transcription, translation and protein folding) can be modelled as a single Poisson process with rate constant $k_{\rm prod}$. The system is symmetric on exchanging A and B. 

\begin{figure}[b]
\begin{center}
\includegraphics[width=0.5\columnwidth,clip=true]{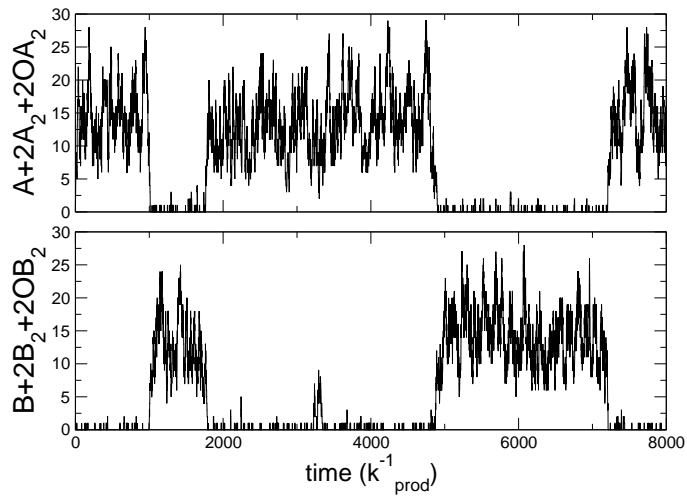}
\caption{Typical simulation trajectory for the model switch, with baseline parameters except for $\mu$ which is replaced by $\mu=0.45k_{\rm prod}$.
The total numbers of A and B molecules fluctuate around two stable states, one rich in A and the other rich in B. Transitions between these
states are rapid, yet infrequent.}
\label{fig:tracks_cg}
\end{center}
\end{figure}

For this model system, bistability has been demonstrated using a mean field analysis and with simulations  \cite{Warren05}. In one stable
state, a large number of A proteins are present; this ensures that the operator $O$ is mostly bound by $\mathrm{A}_2$, keeping $B$ repressed.
Conversely, in the other stable state, B proteins are abundant, so that $O$ is mostly bound by $\mathrm{B}_2$, and $A$ remains repressed.
Previous work has demonstrated that, when simulated stochastically with appropriate parameters, the system  makes occasional random flipping
transitions between these two stable states \cite{Warren05}, as in Figure \ref{fig:tracks_cg}. In our simulations, we use the inverse of the production
rate, $k_{\rm prod}^{-1}$ as the unit of time; typical values for this rate are in the order of $10^{-3}-10^{-4}$s$^{-1}$. 
We assume that the cell volume $V$ remains constant. For simplicity, we use a value $V\!=\!1$,
and define our rate constants in appropriate units. We choose a ``baseline'' set of parameters, in the region of parameter space where the
system has previously been found to be bistable: $k_{\rm f}\!=\!5k_{\rm prod}V$, $k_{\rm b}\!=\!5k_{\rm prod}$ (so
that $K_{\rm D}^d\!=\!k_{\rm d}/k_{\rm f}\!=\!1/V$), $k_{\rm on}\!=\!5k_{\rm
prod}$, $k_{\rm off}\!=\!k_{\rm prod}$ (so that $K_{\rm D}^b\!=\!k_{\rm
off}/k_{\rm on}\!=\!1/(5V)$), $\mu\!=\!0.3k_{\rm prod}$. This model system is loosely based on the bacteriophage lambda genetic switch
\cite{ptashne86}. For the phage lambda proteins cro and cI, assuming diffusion-limited protein-DNA association and using refs
\cite{Darling00} and \cite{McClure83}, $k_{\rm on}\approx 5\!-\!10 k_{\rm prod}$. For cI and cro dimer formation, using refs  \cite{Darling00}
and \cite{Burz94}, $k_{\rm f}\approx50\!-\!100 k_{\rm prod}$. Protein degradation rates are much lower for phage lambda (of the order of
$\mu\approx 0.1\!-\!0.01 k_{\rm prod}$) than for our model system.

Throughout this paper, we represent the number of molecules of chemical species $X$ which is present in the cell by $n_{\rm{X}}$. Later in the paper, we will need to characterize the switching process by an ``order parameter'', which we denote $\lambda$. A natural choice is  the difference between the total numbers of the two proteins in the cell: $\lambda \equiv n_{\rm A}+2n_{\rm
A_2}+2n_{O{\rm A_2}}-(n_{\rm B}+2n_{\rm B_2}+2n_{O{\rm B_2}})$. Figure \ref{fig:tracks_cg} shows $\lambda$ plotted as a function of time for a simulation of this reaction set using the SSA. Bistable behaviour is indeed observed: the system spends most of its
time in one of the two stable states with occasional transitions between states. The average duration of a flipping transition event is much shorter than the average ``waiting time'' between the flipping transitions.

\section{\large{Dynamical coarse-graining: background}}
\label{sec:background}

Dynamical coarse-graining schemes begin by splitting the reaction set into fast and slow reactions, as described in Section
\ref{sec:intro_cg}. The slow reactions are generally simulated using the SSA. The fast reactions are approximated in ways that differ for
different methods. In this paper, we only consider approaches in which the fast reactions are removed entirely from the reaction set, by
assuming that they relax to a steady state faster than the waiting time between slow reactions. Effective propensities for the slow reactions
are computed as averages over the steady state distribution, obtained from the chemical master equation for the fast reactions.

The key step is the determination of the effective
propensity functions $\overline{a}_j^s(n_s,n_f)$ for the slow reactions in the coarse-grained reaction scheme. These depend on
the copy numbers $n_s$ of the ``slow species'' (those
that are only affected by the slow reactions), and $n_f$ of the ``fast species'' (those that are
affected by both the fast and the slow reactions). The effective propensities are given by
\begin{equation}
\overline{a}_j^s(n_f,n_s) = \sum_{n_f^\prime} P_\infty(n_f^\prime | n_f,
n_s) a_j^s(n_f^\prime,n_s),
\label{eq:a_s}
\end{equation}
where $a_j^s$ denotes the propensity function for a given slow
reaction $j$ and $P_\infty(n_f^\prime | n_f, n_s)$ is the probability
of obtaining a given copy number $n_f^\prime$ for the fast species, at
the end of a very long simulation of the fast reaction set only,
starting from state space point ($n_f, n_s$). These effective
propensities are designed to give the same flux along the slow
reaction channel, on average, as in the full system.  In essence, it
is assumed that in the full simulation, for any configuration $n_s$ of
the slow species, the fast species reach a ``quasi-steady state'',
with probability distribution function $P(n_{f} | n_s)$, on a
timescale that is much faster than the waiting time before the next
slow reaction.  In practice, the effective propensity functions in
Eq. (\ref{eq:a_s}) can be obtained by performing short SSA simulations
of the fast reactions at fixed copy numbers of the slow species
\cite{E05,Salis05}. Alternatively, one may solve the chemical master
equation for the fast reactions analytically or numerically
\cite{Bundschuh03,Rao03,Cao05}, which is the approach we employ here.

It is important to discuss the definition of the ``fast variables'' ($n_f$ in Eq. (\ref{eq:a_s})) and ``slow variables'' ($n_s$)
\cite{Bundschuh03,Cao05,Warren06}. In the work of Cao {\em{et al.}} \cite{Cao05}, the slow variables are the copy numbers of those species
which are unaffected by the fast reactions, while the  fast variables can be changed by both fast and slow reactions. During the
coarse-grained simulation, only the slow reactions are simulated, and one has the option of propagating in time either  both the fast and the
slow variables or just the slow variables. Bundschuh {\em{et al.}} \cite{Bundschuh03} describe a slightly different method for eliminating
both the fast reactions and the fast variables from the simulation scheme. Here, as in the work of Cao {\em{et al.}}, the fast variables are
the copy numbers of all species which are affected by the fast reactions. The slow variables, however,  are made up of the copy numbers of
species which are unchanged by the fast reactions, as well as {\em{combinations}} of the fast variables. These combinations are chosen so
that they are unchanged by the fast reactions. For example, for a fast reaction set  $2{\rm A} \rightleftharpoons {\rm
  A_2}$, an appropriate slow variable would be ${{n}}_{\check {\rm{A}}} = n_{\rm A} + 2n_{\rm A_2}$. The original slow reaction set is then
  rewritten in terms of these new slow variables. This eliminates all the fast species from the slow reaction set and the simulation proceeds
  by simulating only the slow variables. This is the strategy which we adopt in this paper.

\section{\large{Coarse-graining for the model genetic switch}}\label{sec:coarse-grain}

Stochastic simulations of the model genetic switch
have characteristic features that depend on the timescale over which
we observe the simulation. In a timeframe of
about $0.2k_{\rm prod}^{-1}$, we will observe mainly protein-protein association   and dissociation events (typical timescale $[n_{\mathrm
A}(n_{\mathrm A}\!-\!1)k_f]^{-1}$ and $[n_{\mathrm
  A_2}k_b]^{-1}$ respectively), as well as protein-operator association and dissociation (typical
timescale $[n_{\mathrm A_2}k_{\mathrm{on}}]^{-1}$ and $[k_{\mathrm{off}}]^{-1}$). If we extend our observation
``window'' to a timeframe of about $4k_{\mathrm{prod}}^{-1}$, we observe also protein production and degradation (typical timescale
$[k_{\mathrm{prod}}]^{-1}$) and $\mu^{-1} \!\approx\! 3k_{\mathrm{prod}}^{-1}$). In a much longer timeframe, we observe flipping
events between the two stable states. It is these flipping
events that are the phenomenon of interest - yet
 for each ``interesting'' switch flipping event, very many
``less interesting'' association and dissociation events need to be
simulated. This is an example of timescale separation, which we seek to overcome by coarse-graining - eliminating the  protein-protein and/or
protein-DNA association and dissociation reactions  from the simulation scheme.

To coarse-grain the model genetic switch, we divide the full reaction set (Table \ref{tab:reactions}) into fast and slow reactions. We will
consider three cases: (i) protein-protein association and dissociation reactions (a in Table \ref{tab:reactions}) are fast,
(ii): protein-DNA association and dissociation reactions (b in Table \ref{tab:reactions}) are fast, and (iii): both reactions
(a) and
(b) are fast. For each of these cases, we define fast and slow variables. The fast variables are the copy numbers $n_f$ of species which are
affected by the fast reactions. The slow variables  $n_s$ are either the copy numbers of species unaffected by the fast reactions, or linear
combinations of fast variables which are unchanged by any of the fast reactions - e.g. ${{n}}_{\check {\rm{A}}} =
n_{\rm A} + 2n_{\rm A_2}$. To accompany these slow variables, we define ``slow species''. These may represent either single chemical species,
or combinations of species. For example, the species $\check {\rm{A}}$ represents an ${\rm{A}}$ molecule which is in either monomer or dimer
form. We then rewrite the slow reaction set in terms of the new slow species, and
carry out a simulation using Gillespie's SSA, with effective propensities given by Eq. (\ref{eq:a_s}). For this system, only the first moment
of $P_\infty(n_f^\prime | n_f,n_s)$ is required. This can be obtained by solving the chemical master equation for the fast reactions at fixed
value of the slow variables. Alternatively, one may approximate to the macroscopic rate equations for the fast reactions. A summary of the
various coarse-graining methods used in the paper is given 
in Table \ref{tab:approx}. Table \ref{tab:approx} also lists the coarse-grained reaction sets, and gives formulae for the effective
propensity functions. The notation $\langle\mathrm X\rangle^{\mathrm{METHOD}}_{\mathrm{SLOW\;VARIABLES}}$ denotes the first moment (average)
of the steady state probability distribution function $P_\infty(n_f^\prime | n_f,n_s)$, the superscript denoting whether the master equation
or macroscopic rate equation is used to find the average, and the subscript indicating which slow variables the average depends upon.


\begin{table*}
\begin{center}
\begin{tabular}{|c|c|c|c|c|c|}
\hline
Name & Reaction & Propensity & Method & Coarse-grained & Definition\\
 &  &  &  & variable & \\
\hline\hline
EO & $\emptyset \to \mathrm{A}$ & $k_{\rm prod}\langle n_{O}+n_{O\rm A_2}\rangle_{\mathrm{\hat{A}_2},\mathrm{\hat{B}_2}}^{\mathrm{RE}}$ & Rate Equation &
$\mathrm{\hat{A}_2}$ & $n_{\mathrm{\hat{A}_2}} = n_{\mathrm{A}_2}+n_{O\mathrm{A}_2}$ \\
  & $\emptyset \to \mathrm{B}$ &  $k_{\rm prod}\langle n_{O}+n_{O\rm B_2}\rangle_{\mathrm{\hat{A}_2},\mathrm{\hat{B}_2}}^{\mathrm{RE}}$ & & $\mathrm{\hat{B}_2}$ & $n_{\mathrm{\hat{B}_2}} =
  n_{\mathrm{B}_2}+n_{O\mathrm{B}_2}$ \\
  & $\mathrm{A}+\mathrm{A}\rightleftharpoons \mathrm{\hat{A}_2}$ & $k_{\rm on} n_{\rm A}(n_{\rm A}\!-\!1),\; k_{\rm off}\langle n_{\rm
  A_2}\rangle_{\mathrm{\hat{A}_2},\mathrm{\hat{B}_2}}^{\mathrm{RE}}$ & &  &\\
  & $\mathrm{B}+\mathrm{B}\rightleftharpoons \mathrm{\hat{B}_2}$ & $k_{\rm on} n_{\rm B}(n_{\rm B}\!-\!1),\; k_{\rm off}\langle n_{\rm
  B_2}\rangle_{\mathrm{\hat{A}_2},\mathrm{\hat{B}_2}}^{\mathrm{RE}}$ & &  &\\
  & $\mathrm{A} \to \emptyset$ & $\mu n_{\rm A}$ & &  &\\
  & $\mathrm{B} \to \emptyset$ & $\mu n_{\rm B}$ & &  &\\
\hline
ED1 & ${O}+2\check{\mathrm{A}}\rightleftharpoons O\mathrm{A}_2$ & $k_{\rm on}\langle n_{\rm
A_2}\rangle_{\check{\mathrm{A}}}^{\mathrm{RE}},\; k_{\rm off} n_{O\rm
  A_2}$ & Rate Equation & $\check{\mathrm{A}}$ & $n_{\check{\mathrm{A}}} = n_{\mathrm{A}}+2n_{\mathrm{A}_2}$  \\
  & ${O}+2\check{\mathrm{B}}\rightleftharpoons O\mathrm{B}_2$ & $k_{\rm on}\langle n_{\rm B_2}\rangle_{\check{\mathrm{B}}}^{\mathrm{RE}}, \; k_{\rm off} n_{O\rm
  B_2}$ & &  $\check{\mathrm{B}}$ & $n_{\check{\mathrm{B}}} =n_{\mathrm{B}}+2n_{\mathrm{B}_2}$ \\
  & ${O} \to {O} +\check{\mathrm{A}}$ & $k_{\rm prod} n_{O}$   & &  &\\
  & ${O} \to {O} +\check{\mathrm{B}}$ & $k_{\rm prod} n_{O}$  & &  &\\
  & $O\mathrm{A}_2 \to O\mathrm{A}_2 + \check{\mathrm{A}}$ & $k_{\rm prod} n_{O\rm A_2}$ & &  &\\
  & $O\mathrm{B}_2 \to O\mathrm{B}_2 + \check{\mathrm{B}}$ & $k_{\rm prod} n_{O\rm B_2}$ & &  &\\
  & $\check{\mathrm{A}} \to \emptyset$ & $\mu \langle n_{\rm A}\rangle_{\check{\mathrm{A}}}^{\mathrm{RE}}$ & &  &\\
  & $\check{\mathrm{B}} \to \emptyset$ & $\mu \langle n_{\rm B}\rangle_{\check{\mathrm{B}}}^{\mathrm{RE}}$ & &  &\\
\hline
ED2 & ${O}+2\check{\mathrm{A}}\rightleftharpoons O\mathrm{A}_2$ & $k_{\rm on}\langle n_{\rm
A_2}\rangle_{\check{\mathrm{A}}}^{\mathrm{ME}}, \;k_{\rm off}
n_{O\rm A_2}$ & Master Equation & $\check{\mathrm{A}}$ &$n_{\check{\mathrm{A}}} =n_{\mathrm{A}}+2n_{\mathrm{A}_2}$ \\
  & ${O}+2\check{\mathrm{B}}\rightleftharpoons O\mathrm{B}_2$ & $k_{\rm on}\langle n_{\rm B_2}\rangle_{\check{\mathrm{B}}}^{\mathrm{ME}}, \; k_{\rm off} n_{O\rm
  B_2}$ & & $\check{\mathrm{B}}$ &$n_{\check{\mathrm{B}}} =n_{\mathrm{B}}+2n_{\mathrm{B}_2}$\\
  & ${O} \to {O} +\check{\mathrm{A}}$ & $k_{\rm prod} n_{O}$ & &  &\\
  & ${O} \to {O} +\check{\mathrm{B}}$ & $k_{\rm prod} n_{O}$ & &  &\\
  & $O\mathrm{A}_2 \to O\mathrm{A}_2 + \check{\mathrm{A}}$ & $k_{\rm prod} n_{O\rm A_2}$ & &  &\\
  & $O\mathrm{B}_2 \to O\mathrm{B}_2 + \check{\mathrm{B}}$ & $k_{\rm prod} n_{O\rm B_2}$ & &  &\\
  & $\check{\mathrm{A}} \to \emptyset$ & $\mu \langle n_{\rm A}\rangle_{\check{\mathrm{A}}}^{\mathrm{ME}}$ & &  &\\
  & $\check{\mathrm{B}} \to \emptyset$ & $\mu \langle n_{\rm B}\rangle_{\check{\mathrm{B}}}^{\mathrm{ME}}$ & &  &\\
\hline
EO-ED1 & $\emptyset \to \tilde{\mathrm{A}}$ & $k_{\rm prod}\langle n_{O}+n_{O\rm
A_2}\rangle_{\tilde{\mathrm{A}},\tilde{\mathrm{B}}}^{\mathrm{RE}}$ & Rate Equation &$\tilde{\mathrm{A}}$ &
$n_{\tilde{\mathrm{A}}} = n_{\mathrm{A}}+2n_{\mathrm{A}_2}+2n_{O\mathrm{A}_2}$\\
  & $\emptyset \to \tilde{\mathrm{B}}$ & $k_{\rm prod}\langle n_{O}+n_{O\rm
  B_2}\rangle_{\tilde{\mathrm{A}},\tilde{\mathrm{B}}}^{\mathrm{RE}}$ & & $\tilde{\mathrm{B}}$ &
  $n_{\tilde{\mathrm{B}}} = n_{\mathrm{B}} +2n_{\mathrm{B}_2}+2n_{O\mathrm{B}_2}$\\
  & $\tilde{\mathrm{A}} \to \emptyset$ & $\mu \langle n_{\rm A}\rangle_{\tilde{\mathrm{A}},\tilde{\mathrm{B}}}^{\mathrm{RE}}$ & &  &\\
  & $\tilde{\mathrm{B}} \to \emptyset$ & $\mu \langle n_{\rm B}\rangle_{\tilde{\mathrm{A}},\tilde{\mathrm{B}}}^{\mathrm{RE}}$ & &  &\\
\hline
EO-ED2 & $\emptyset \to \tilde{\mathrm{A}}$ & $k_{\rm prod}\langle n_{O}+n_{O\rm
A_2}\rangle_{\tilde{\mathrm{A}},\tilde{\mathrm{B}}}^{\mathrm{ME}}$ & Master Equation &$\tilde{\mathrm{A}}$ &
$n_{\tilde{\mathrm{A}}} = n_{\mathrm{A}}+2n_{\mathrm{A}_2}+2n_{O\mathrm{A}_2}$ \\
  & $\emptyset \to \tilde{\mathrm{B}}$ & $k_{\rm prod}\langle  n_{O}+n_{O\rm
  B_2}\rangle_{\tilde{\mathrm{A}},\tilde{\mathrm{B}}}^{\mathrm{ME}}$ & & $\tilde{\mathrm{B}} $ &
  $n_{\tilde{\mathrm{B}}} = n_{\mathrm{B}} +2n_{\mathrm{B}_2}+2n_{O\mathrm{B}_2}$\\
  & $\tilde{\mathrm{A}} \to \emptyset$ & $\mu \langle n_{\rm A}\rangle_{\tilde{\mathrm{A}},\tilde{\mathrm{B}}}^{\mathrm{ME}} $ & &  &\\
  & $\tilde{\mathrm{B}} \to \emptyset$ & $\mu \langle n_{\rm B}\rangle_{\tilde{\mathrm{A}},\tilde{\mathrm{B}}}^{\mathrm{ME}} $ & &  &\\
\hline
\end{tabular}
\end{center}
\caption{Summary of coarse-graining schemes for the original reaction set
(Table \ref{tab:reactions}): eliminating operator binding (EO), eliminating dimerisation reactions using the Macroscopic Rate Equation (ED1) or the Master
Equation (ED2), eliminating both dimerisation and operator binding using the Macroscopic Rate Equation (EO-ED1) or the Master Equation
(EO-ED2). For each coarse-graining scheme, the coarse-grained reaction set is indicated together with the propensity function for each reaction. We also give definitions of the new slow variables for each scheme.
\label{tab:approx}}
\end{table*}


\subsection{\large{Coarse-graining protein-DNA binding}}\label{subsec:av_operator}
We first remove the protein-DNA association and dissociation reactions
\begin{equation}
\begin{array}{lcl}\label{eq:MM_eliminated}
{O} + \mathrm{A}_2 \rightleftharpoons {O}\mathrm{A}_2 & \quad & {O} + \mathrm{B}_2 \rightleftharpoons
{O}\mathrm{B}_2
\end{array}
\end{equation}
from the original reaction scheme (Table \ref{tab:reactions}). We denote this approach ``Eliminating Operator state fluctuations (EO)''. In our coarse-grained simulation, the system will still experience fluctuations due to protein-protein association and dissociation, protein production and protein decay, but not those due to the binding and unbinding of molecules to the DNA.

The ``fast species'', which are affected by reactions (\ref{eq:MM_eliminated}), are ${\rm A_2}$,  ${\rm B_2}$, $O{\rm A_2}$, $O{\rm B_2}$ and ${O}$. The ``slow species'' are ${\rm A}$, ${\rm B}$, $\mathrm{\hat{A}_2}$ and $\mathrm{\hat{B}_2}$, where ${\rm A}$ and ${\rm B}$ are simply the protein monomers - these are unchanged by the fast reactions (\ref{eq:MM_eliminated}) - and $\mathrm{\hat{A}_2}$ and $\mathrm{\hat{B}_2}$ are new species, such that:
\begin{eqnarray}
n_{\mathrm{\hat{A}_2}}&=&n_{\rm A_2}+n_{O\rm A_2} \\
\nonumber n_{\mathrm{\hat{B}_2}}&=&n_{\rm B_2}+n_{O\rm B_2}
\end{eqnarray}
$n_{\mathrm{\hat{A}_2}}$ and $n_{\mathrm{\hat{B}_2}}$ are simply the total numbers of dimers in the system - including both free and DNA-bound dimers. The new, coarse-grained, reaction set is given 
in Table \ref{tab:approx}, together with the effective propensities. As the operator $O$ has been removed from the scheme, protein production
has become a simple birth process, in which a monomer appears from ``nowhere''. The propensity for this birth of a monomer (say ${\rm A}$)
takes into account the ``lost'' reactions - it reflects the probability, in the full reaction scheme, of finding the promoter ${O}$ in one of
the states $O$ and $O\mathrm{A_2}$ that are able to produce ${\rm A}$. The protein-protein interactions ((a) in Table \ref{tab:reactions}) have been
changed to reflect the fact that free dimers ${\rm{A_2}}$ have been replaced by the new species $\mathrm{\hat{A}_2}$. Two monomers can now
reversibly associate to generate a molecule of  $\mathrm{\hat{A}_2}$, while the reaction representing dissociation of free dimers to monomers
has an effective propensity that depends on the average number of free dimers that would be present in the full reaction scheme, for a given
value $n_{\mathrm{\hat{A}_2}}$ of total dimers. Protein degradation ((e) in Table \ref{tab:reactions}) remains unchanged since these reactions affect only monomers.

To evaluate the effective propensities listed in Table \ref{tab:approx}, we require the averages $\langle n_{O}\!+\!n_{O\rm
A_2}\rangle_{\mathrm{\hat{A}_2},\mathrm{\hat{B}_2}}$, $\langle n_{O}\!+\!n_{O\rm
B_2}\rangle_{\mathrm{\hat{A}_2},\mathrm{\hat{B}_2}}$,$\langle n_{\rm A_2}\rangle_{\mathrm{\hat{A}_2},\mathrm{\hat{B}_2}}$ and  $\langle
n_{\rm B_2}\rangle_{\mathrm{\hat{A}_2},\mathrm{\hat{B}_2}}$. These  depend on both slow species $\mathrm{\hat{A}_2}$ and
$\mathrm{\hat{B}_2}$, because the two operator binding reactions are coupled. This arises from the competition between $\mathrm{A}_2$ and
$\mathrm{B}_2$ for the same
binding site. In this particular case, as shown in Appendix \ref{app:ME2}, solving the master equation for the fast reactions
(\ref{eq:MM_eliminated}) and approximating them by
the corresponding macroscopic rate equations give the same result, so we consider only the
rate equation approach. 
Solving for the steady state of Eqs. (\ref{eq:MM_eliminated}), we obtain
\begin{eqnarray}\label{test1}
K_{\rm D}^b \langle n_{O\mathrm{A_2}}\rangle_{\mathrm{\hat{A}_2},\mathrm{\hat{B}_2}}^{\mathrm{RE}}&=&\langle n_{{O}}\rangle_{\mathrm{\hat{A}_2},\mathrm{\hat{B}_2}}^{\mathrm{RE}}\cdot \langle n_{\mathrm{A_2}}\rangle_{\mathrm{\hat{A}_2},\mathrm{\hat{B}_2}}^{\mathrm{RE}}\\
\nonumber K_{\rm D}^b \langle n_{O\mathrm{B_2}}\rangle_{\mathrm{\hat{A}_2},\mathrm{\hat{B}_2}}^{\mathrm{RE}}& = &\langle n_{{O}}\rangle_{\mathrm{\hat{A}_2},\mathrm{\hat{B}_2}}^{\mathrm{RE}}\cdot \langle n_{\mathrm{B_2}}\rangle_{\mathrm{\hat{A}_2},\mathrm{\hat{B}_2}}^{\mathrm{RE}}.
\end{eqnarray}
Combining this with the fact that in our scheme there is only one DNA copy: 
\begin{equation}\label{test2}
n_{{O}}+n_{O\mathrm{A_2}}+n_{O\mathrm{B_2}}=1
\end{equation}
we obtain 
\begin{eqnarray}\label{eq:prop_eff_B}
a_{\rm A,eff} &=& k_{\rm prod}\langle n_{O}+n_{O\rm
A_2}\rangle_{\mathrm{\hat{A}_2},\mathrm{\hat{B}_2}}^{\mathrm{RE}}\\\nonumber &=& k_{\rm prod}
\frac{1+(K_{\rm D}^b)^{-1}\langle n_{\mathrm A_2}\rangle_{\mathrm{\hat{A}_2},\mathrm{\hat{B}_2}}^{\mathrm{RE}}}
{1+(K_{\rm D}^b)^{-1}\left (\langle n_{\mathrm A_2}\rangle_{\mathrm{\hat{A}_2},\mathrm{\hat{B}_2}}^{\mathrm{RE}}
+\langle n_{\mathrm B_2}\rangle_{\mathrm{\hat{A}_2},\mathrm{\hat{B}_2}}^{\mathrm{RE}}\right )},
\end{eqnarray}
and similarly for $a_{\rm B,eff}$. To find $\langle n_{\mathrm A_2}\rangle_{\mathrm{\hat{A}_2},\mathrm{\hat{B}_2}}^{\mathrm{RE}}$  and  $\langle n_{\mathrm B_2}\rangle_{\mathrm{\hat{A}_2},\mathrm{\hat{B}_2}}^{\mathrm{RE}}$ in Eq. (\ref{eq:prop_eff_B}), we combine relations (\ref{test1}) and (\ref{test2}) with
\begin{eqnarray}
n_{\mathrm{\hat{A}_2}}&=&n_{\rm A_2}+n_{O\rm A_2}\\
\nonumber n_{\mathrm{\hat{B}_2}}&=&n_{\mathrm B_2}+n_{O\mathrm B_2}.
\end{eqnarray}
Numerical solution techniques are required here, and we have used the Newton-Raphson method \cite{NumRec}.

\subsection{\large{Coarse-graining protein-protein binding}}\label{subsec:av_dimerisation}

We now remove instead the protein-protein association and dissociation reactions
\begin{eqnarray}\label{ed1:elim}
\mathrm{A}+\mathrm{A} &\rightleftharpoons \mathrm{A}_2\\ 
\nonumber \mathrm{B}+\mathrm{B} &\rightleftharpoons \mathrm{B}_2
\end{eqnarray}
from our original reaction scheme (Table \ref{tab:reactions}). We denote this approach ``Eliminating dimerisation (ED)''. These interactions
are particularly attractive candidates for coarse-graining, since they tend to consume a large fraction of the computational effort when
there are significant numbers of free monomers and dimers in the system. 

The ``fast'' species - those whose number is affected by reactions (\ref{ed1:elim}) - are ${\rm A}$, ${\rm A_2}$, ${\rm B}$ and ${\rm B_2}$.
The ``slow species'', which will remain in our coarse-grained reaction scheme, are ${O}$, ${O\rm A_2}$, ${O\rm B_2}$ - species from the
original scheme which are not affected by reactions (\ref{ed1:elim}) - together with two new species, $\check{\mathrm{A}}$ and
$\check{\mathrm{B}}$, defined by:
\begin{eqnarray}\label{ed1:defs}
n_{\check{\mathrm{A}}} & \equiv n_{{\rm A}}+2n_{{\rm A_2}}\\
\nonumber n_{\check{\mathrm{B}}} & \equiv n_{{\rm B}}+2n_{{\rm B_2}}
\end{eqnarray}
 These new species  $\check{\mathrm{A}}$ and $\check{\mathrm{B}}$ are combinations of the fast species whose number remains unchanged by the
 fast reactions ((a) in Table \ref{tab:reactions}).  The new, coarse-grained, reaction set with the corresponding propensity functions, 
 is given in Table \ref{tab:approx}. The protein production reactions ((c) and (d) in Table \ref{tab:reactions}) now produce
 the new species $\check{\mathrm{A}}$ and $\check{\mathrm{B}}$. In the original reaction set (Table \ref{tab:reactions}), the protein
 degradation reactions (e) affected only monomers. The corresponding reaction in the new reaction set removes a
 molecule of the new species $\check{\mathrm{A}}$ and $\check{\mathrm{B}}$ from the system, but with an effective propensity that depends on
 the average number of monomers that would be obtained by a simulation of the fast reactions, at fixed  $n_{\check{\mathrm{A}}}$ or
 $n_{\check{\mathrm{B}}}$. Similarly, reactions (b) in the original set, corresponding to the association and dissociation of dimers with the DNA, have been replaced by the association/dissociation of two units of $\check{\mathrm{A}}$ or $\check{\mathrm{B}}$  to $O$, with an effective propensity proportional to the average number of dimers given by the fast reaction set for fixed $n_{\check{\mathrm{A}}}$ or $n_{\check{\mathrm{B}}}$. Here, the averages required for the effective propensity functions depend on only one slow variable - either $\check{\mathrm{A}}$ or $\check{\mathrm{B}}$ but not both - in contrast to method  EO, where the averages depend on both slow variables. This is because the two reactions (\ref{ed1:elim}) are not coupled to each other: dimerisation of {\rm A} has no direct effect on the dimerisation propensity of {\rm B} and vice versa.   

We shall test two alternative approaches to the computation of the averages $\langle n_{\mathrm A}\rangle_{\check{\mathrm{A}}}$, $\langle n_{\mathrm A_2}\rangle_{\check{\mathrm{A}}}$, $\langle n_{\mathrm B}\rangle_{\check{\mathrm{B}}}$ and $\langle n_{\mathrm B_2}\rangle_{\check{\mathrm{B}}}$ 
in Table \ref{tab:approx}. In the first approach, which we denote ED1, we make the approximation that these averages correspond to the steady state solutions of the macroscopic rate equations corresponding to (\ref{ed1:elim}): 
\begin{eqnarray}\label{eq:dim_eq}
k_{b}\langle n_{\mathrm A_2}\rangle_{\check{\mathrm{A}}}-k_{\rm f}\langle n_{\mathrm A}\rangle_{\check{\mathrm{A}}}^2 & =0\\
\nonumber k_{b}\langle n_{\mathrm B_2}\rangle_{\check{\mathrm{B}}}-k_{\rm f}\langle n_{\mathrm B}\rangle_{\check{\mathrm{B}}}^2 & =0
\end{eqnarray}
so that 
\begin{eqnarray}\label{ed1:rel1}
K_{\rm D}^d \langle n_{\mathrm A_2}\rangle_{\check{\mathrm{A}}} & =\langle n_{\mathrm A}\rangle_{\check{\mathrm{A}}}^2\\
\nonumber K_{\rm D}^d \langle n_{\mathrm B_2}\rangle_{\check{\mathrm{B}}} & =\langle n_{\mathrm B}\rangle_{\check{\mathrm{B}}}^2.
\end{eqnarray}
Relations (\ref{ed1:rel1}) can be used together with the definitions (\ref{ed1:defs}) to give
\begin{eqnarray}\label{ed1:resol}
\langle n_{\mathrm A}\rangle_{\check{\mathrm{A}}}^{\mathrm{RE}}=K_{\rm D}^d \left(\sqrt{8 n_{\check{\mathrm{A}}}/K_{\rm D}^d +1}-1
\right)/4\\
\nonumber \langle n_{\mathrm B}\rangle_{\check{\mathrm{B}}}^{\mathrm{RE}}=K_{\rm D}^d \left(\sqrt{8 n_{\check{\mathrm{B}}}/K_{\rm D}^d +1}-1
\right)/4.
\end{eqnarray}
The average numbers of dimers $\langle n_{\mathrm
  A_2}\rangle_{\check{\mathrm{A}}}$ and $\langle n_{\mathrm
  B_2}\rangle_{\check{\mathrm{B}}}$ are given in this approximation by
combining (\ref{ed1:resol}) with (\ref{ed1:rel1}). Method ED1 is
expected to give incorrect results when $n_{\check{\mathrm{A}}}$ or
$n_{\check{\mathrm{B}}}$ is small, since the macroscopic rate equation
approximation breaks down in this limit. This is expected to be a
serious problem, because for our genetic switch model, both
$n_{\check{\mathrm{A}}}$ and $n_{\check{\mathrm{B}}}$ will be small at
the crucial moments when the switch is in the process of flipping
between the two steady states \cite{Warren05}. Alternatively, one may solve the master equation corresponding to the
eliminated reactions (\ref{ed1:elim}) to compute the averages. We
denote this approach ED2. Numerical solution of this master equation,
as described in Appendix \ref{app:ME1}, results in the probability
distribution functions $p(n_{\mathrm{A}}|n_{\check{\mathrm{A}}})$,
$p(n_{\mathrm{A_2}}|n_{\check{\mathrm{A}}})$,
$p(n_{\mathrm{B}}|n_{\check{\mathrm{B}}})$ and
$p(n_{\mathrm{B_2}}|n_{\check{\mathrm{B}}})$ for the fast variables,
for given values of $\check{\mathrm{A}}$ and
$\check{\mathrm{B}}$. These can be used to find $\langle n_{\mathrm
  A}\rangle_{\check{\mathrm{A}}}^{\mathrm{ME}}$, $\langle n_{\mathrm
  A_2}\rangle_{\check{\mathrm{A}}}^{\mathrm{ME}}$, $\langle n_{\mathrm
  B}\rangle_{\check{\mathrm{B}}}^{\mathrm{ME}}$ and $\langle
n_{\mathrm B_2}\rangle_{\check{\mathrm{B}}}^{\mathrm{ME}}$. These
averages are then substituted into the expressions given in Table
\ref{tab:approx} to obtain effective propensities for the
coarse-grained SSA simulation. We note that the effective propensity
functions for methods ED1 and ED2 in Table \ref{tab:approx} are
identical.  The only difference between the two schemes is the way in
which the required averages are obtained: using a
macroscopic rate equation approximation (ED1) or by
numerical solution of the master equation (ED2).

\subsection{\large{Coarse-graining protein-DNA and protein-protein binding}}
We now  eliminate both protein-DNA interactions [Eq.(\ref{eq:MM_eliminated})] and protein-protein interactions  [Eq.(\ref{ed1:elim})]. We will be left with a coarse-grained scheme in which the only fluctuations are due to protein production and degradation. Our ``fast reactions'' are then (\ref{eq:MM_eliminated}) and (\ref{ed1:elim}), and our ``fast species'', whose number is changed by the fast reactions, are in fact {\em{all}} the species in the original scheme: ${O}$, $O\rm{A_2}$, $O\rm{B_2}$, $\rm{A}$,  $\rm{A_2}$, $\rm{B}$ and $\rm{B_2}$. Our only slow species, $\tilde{\mathrm A}$ and  $\tilde{\mathrm B}$,  are then combinations of the fast species whose number is unchanged in any of the fast reactions:
\begin{eqnarray}\label{eoed:defs}
n_{\tilde{\mathrm A}} & \equiv n_{\rm A}+2n_{\rm A_2}+2n_{O\rm A_2}\\
\nonumber n_{\tilde{\mathrm B}} & \equiv n_{\rm B}+2n_{\rm B_2}+2n_{O\rm B_2}.
\end{eqnarray}
$n_{\tilde{\mathrm A}}$ and $n_{\tilde{\mathrm B}}$ correspond to the total number of A and B molecules in the system. On removal of reactions (\ref{eq:MM_eliminated}) and (\ref{ed1:elim}), 
our coarse-grained reaction scheme, given in Table \ref{tab:sim} under the labels EO-ED1 and EO-ED2, consists simply of a pair of birth-death processes for species $\tilde{\mathrm A}$ and $\tilde{\mathrm B}$. The effects of the ``lost'' fast reactions are incorporated via effective rate constants that account for the average number of the relevant fast species expected in a simulation of the fast reaction set, for fixed numbers of the slow species. As in the EO coarse-graining scheme, but not in the ED schemes, the averages here depend upon both slow species, since the fast reactions for A and B are coupled by the shared DNA binding sites. 

As for the ED schemes, we consider two alternative ways of obtaining the necessary averages $\langle n_{
O}\rangle_{\tilde{\mathrm{A}},\tilde{\mathrm{B}}}$, $\langle n_{\rm A}\rangle_{\tilde{\mathrm{A}},\tilde{\mathrm{B}}}$, $\langle n_{\rm
A_2}\rangle_{\tilde{\mathrm{A}},\tilde{\mathrm{B}}}$,  $\langle n_{O\rm A_2}\rangle_{\tilde{\mathrm{A}},\tilde{\mathrm{B}}}$, $\langle n_{\rm
B}\rangle_{\tilde{\mathrm{A}},\tilde{\mathrm{B}}}$,  $\langle n_{\rm B_2}\rangle_{\tilde{\mathrm{A}},\tilde{\mathrm{B}}}$ and  $\langle
n_{O\rm B_2}\rangle_{\tilde{\mathrm{A}},\tilde{\mathrm{B}}}$. Firstly, in approach EO-ED1, we approximate these averages by the steady state
solutions of the macroscopic rate equations corresponding to the fast reactions (\ref{eq:MM_eliminated}) and (\ref{ed1:elim}). Following the
same steps as in the previous two sections (applying equations (\ref{test1}) and (\ref{eq:dim_eq})), we arrive at 
\begin{eqnarray}
\langle n_{\tilde{\mathrm A}}\rangle_{\tilde{\mathrm{A}},\tilde{\mathrm{B}}}^{\mathrm{RE}} & = & \langle n_{\rm A}\rangle_{\tilde{\mathrm{A}},\tilde{\mathrm{B}}}^{\mathrm{RE}}+2(K_{\rm D}^d)^{-1}\left(\langle n_{\rm A}\rangle_{\tilde{\mathrm{A}},\tilde{\mathrm{B}}}^{\mathrm{RE}}\right)^2+\\\nonumber &&\frac{2(K_{\rm D}^d)^{-1}(K_{\rm D}^b)^{-1} (\langle n_{\rm A}\rangle_{\tilde{\mathrm{A}},\tilde{\mathrm{B}}}^{\mathrm{RE}})^2}{1+(K_{\rm D}^dK_{\rm D}^b)^{-1}\left[(\langle n_{\rm A}\rangle_{\tilde{\mathrm{A}},\tilde{\mathrm{B}}}^{\mathrm{RE}})^2+(\langle n_{\rm B}\rangle_{\tilde{\mathrm{A}},\tilde{\mathrm{B}}}^{\mathrm{RE}})^2\right]}\\ \nonumber
\langle n_{\tilde{\mathrm B}}\rangle_{\tilde{\mathrm{A}},\tilde{\mathrm{B}}}^{\mathrm{RE}} & = & \langle n_{\rm B}\rangle_{\tilde{\mathrm{A}},\tilde{\mathrm{B}}}^{\mathrm{RE}}+2(K_{\rm D}^d)^{-1}\left(\langle n_{\rm B}\rangle_{\tilde{\mathrm{A}},\tilde{\mathrm{B}}}^{\mathrm{RE}}\right)^2+\\\nonumber &&\frac{2(K_{\rm D}^d)^{-1}(K_{\rm D}^b)^{-1}(\langle n_{\rm B}\rangle_{\tilde{\mathrm{A}},\tilde{\mathrm{B}}}^{\mathrm{RE}})^2}{
1+(K_{\rm D}^dK_{\rm D}^b)^{-1}\left[(\langle n_{\rm A}\rangle_{\tilde{\mathrm{A}},\tilde{\mathrm{B}}}^{\mathrm{RE}})^2+(\langle n_{\rm
B}\rangle_{\tilde{\mathrm{A}},\tilde{\mathrm{B}}}^{\mathrm{RE}})^2\right]}, \nonumber
\end{eqnarray}
which can be combined with relations (\ref{eoed:defs}) and inverted numerically \cite{NumRec} to give $\langle n_{\rm A}\rangle_{\tilde{\mathrm{A}},\tilde{\mathrm{B}}}^{\mathrm{RE}}$ and $\langle n_{\rm B}\rangle_{\tilde{\mathrm{A}},\tilde{\mathrm{B}}}^{\mathrm{RE}}$. 
The other averages required in Table \ref{tab:approx} are obtained using relations  (\ref{test1}), (\ref{test2}) and (\ref{ed1:rel1}). Approach EO-ED1 is approximate, since it assumes that the average numbers of molecules that would be produced by long stochastic simulations of the fast reaction set are given by the steady-state solutions of the corresponding macroscopic rate equations. This approximation is avoided in approach EO-ED2, in which we calculate the averages of the fast variables A, ${\rm A}_2$, ${O\rm A}_2$, B,
${\rm B}_2$, ${O\rm B}_2$ and $O$ from the master equation corresponding
to the coupled reactions (\ref{eq:MM_eliminated}) and
(\ref{ed1:elim}). This is difficult to do directly (as in scheme ED2),
so we use simulations. We carry out a series of short
preliminary simulations, using the SSA, of reactions
(\ref{eq:MM_eliminated}) and (\ref{ed1:elim}), for fixed values of
$n_{\tilde{\mathrm A}}$ and $n_{\tilde{\mathrm B}}$. The reaction scheme for these preliminary simulations is given in Table \ref{tab:sim}.


\begin{table*}[t]
\begin{tabular}{cccc}
$\quad\qquad\mathrm{Reaction}\qquad\quad$ &$\qquad\quad\mathrm{Propensity}\qquad\quad$ & $\qquad\quad\mathrm{Reaction}\qquad\quad$ & $\qquad\quad\mathrm{Propensity}\qquad\quad$\\
\hline
$\mathrm{A}+\mathrm{A} \rightleftharpoons \mathrm{A}_2$ & $k_{\rm f}n_{\mathrm{A}}(n_{\mathrm{A}}\!-\!1),\:k_{\rm b}n_{\mathrm{A}_2}$
& $\mathrm{B}+\mathrm{B} \rightleftharpoons \mathrm{B}_2$  & $k_{\rm f}n_{\mathrm{B}}(n_{\mathrm{B}}\!-\!1),\:k_{\rm
b}n_{\mathrm{B}_2}$ \\
${O} + \mathrm{A}_2 \rightleftharpoons {O}\mathrm{A}_2$ & $k_{\rm on} n_{{O}}n_{\mathrm{A}_2},\:k_{\rm
off}n_{O\mathrm{A}_2}$ & ${O} + \mathrm{B}_2 \rightleftharpoons {O}\mathrm{B}_2 $ & $k_{\rm on}
n_{{O}}n_{\mathrm{B}_2},\:k_{\rm off}n_{O\mathrm{B}_2}$
\end{tabular}
\caption{Reaction scheme for the preliminary simulations to
  compute the effective propensity functions given in
  Eqs. (\ref{eq:a_eff}) and (\ref{eq:mu_eff}), for scheme EO-ED2.\label{tab:sim}}
\end{table*}


From these, we compute the averages required for the effective propensities 
\begin{equation}
\label{eq:a_eff}
a_{\rm{A,eff}}=k_{\rm{prod}}\left (\langle n_{O} + n_{O\rm A_2}\rangle_{\tilde{\mathrm{A}},\tilde{\mathrm{B}}}^{\mathrm{ME}}\right )
\end{equation}
and
\begin{equation}
\label{eq:mu_eff}
\mu_{\rm{A,eff}}=\mu \langle n_{\rm A}\rangle_{\tilde{\mathrm{A}},\tilde{\mathrm{B}}}^{\mathrm{ME}}.
\end{equation}
These propensities are stored in a lookup table which is referred to during the coarse-grained simulations of the slow variables.

\section{\large{Results}}
\label{sec:Results}
We now assess the performance of the various coarse-graining approaches, in terms of how accurately they reproduce the behaviour of the full system, and how much they speed up the simulations. Key features of the behaviour of this system are its bimodal steady state probability distribution function and its spontaneous flips between the two stable states. We assess how well the coarse-grained simulations reproduce the bimodal  probability distribution for the difference  $\lambda$ in total number between the A and B proteins:
\begin{equation}\label{Eq:lambda}
\lambda=n_{\rm A}+2n_{\rm A_2}+2n_{O{\rm A_2}}-(n_{\rm B}+2n_{\rm B_2}+2n_{O{\rm B_2}})
\end{equation}
We also test how well the various schemes reproduce the rate of fluctuation-induced switching between the A and B-rich states, which we
measure using the Forward Flux Sampling (FFS) rare event simulation method \cite{Allen05,Allen06_1,Allen06_2}.  We compare our results to SSA
simulations of the full reaction set (Table \ref{tab:reactions}) which we denote ORN (``Original Reaction Network''). The coarse-graining
schemes are based on the assumption that the ``fast'' reactions are indeed fast compared to the slow reactions. We therefore expect their
accuracy to improve as the rate constants for the ``fast'' reactions increase. We will see that the accuracy also depends on which reactions
we choose to be ``fast'' and how we compute the averages needed for the propensity functions. We end with a discussion on the computational
performances of the different methods.

\subsection{\large{Steady-state probability distribution}}\label{sec:eq_results}
\begin{figure}[t]
\begin{center}
\includegraphics[width=0.5\columnwidth,clip=true]{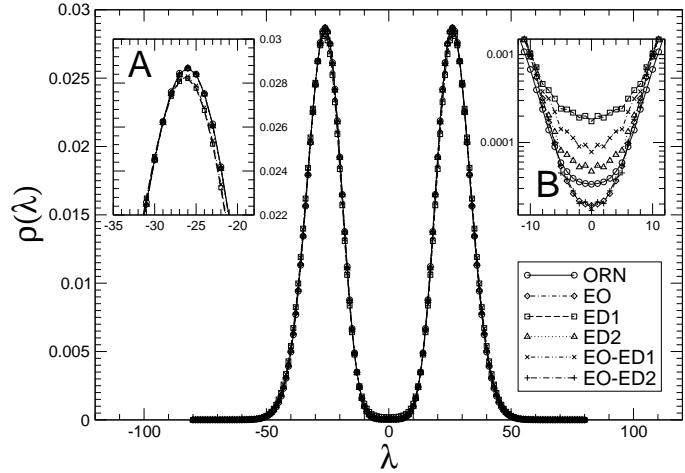}
\caption{Probability distribution $\rho(\lambda)$ as a function of the order parameter $\lambda$.  Inset A zooms in on the left peaks and shows how all
the methods are able to reproduce the positions and heights of the peaks. Inset B shows, on a logarithmic scale, deviations between the different coarse-graining
schemes and the original network for the region around $\lambda\!=\!0$. FFS was used to sample $\rho(\lambda)$ in this region.}
\label{fig:Eq_cg}
\end{center}
\end{figure}

We now compute the steady-state probability distribution function $\rho(\lambda)$ for the difference $\lambda$ between the total numbers of A
and B proteins, as given by Eq. (\ref{Eq:lambda}). We expect $\rho(\lambda)$ to have two peaks around the known stable steady states 
$\lambda\!=\!\pm 27$ (see ref. \cite{Warren05}), and a ``valley'' around the unstable steady state $\lambda\!=\!0$. To compute $\rho(\lambda)$, we carry out a long SSA simulation, during which we compile a histogram for the probability of finding 
the system at each $\lambda$ value. 
This procedure is repeated for all the coarse-grained simulation schemes in Table \ref{tab:approx}. However, it is hard to achieve good
sampling of $\rho(\lambda)$ in the ``valley'' region close to $\lambda=0$, where the system is very unlikely to be found. In this region we
use the FFS method to compute $\rho(\lambda)$ more accurately \cite{Valeriani07}. This method will be described  briefly in the following
section and in Appendix \ref{app:FFS}. Its use for computing steady state probability distributions is described in Ref.\cite{Valeriani07}. Results are shown in Figure \ref{fig:Eq_cg}, for the ``baseline'' parameter set given in section
\ref{sec:GS_cg}. As expected, $\rho(\lambda)$ is clearly bimodal and shows symmetric peaks flanking a ``valley'' at $\lambda\!=\!0$. The
locations of the peaks and valley correspond to the stable and unstable solutions of a mean field analysis \cite{Warren05} of the switch.
Comparing the results for the different coarse-graining schemes in Figure  \ref{fig:Eq_cg}, we see that they all appear to reproduce $\rho(\lambda)$ quite well, giving the correct position, height and width of the peaks. Inset A magnifies the left probability peak, showing that the only methods displaying a small systematic error are
ED1 and EO-ED1, {\em{i.e.}} the coarse-graining schemes relying on the
solutions of the Macroscopic Rate Equation. In general, we can conclude that  all the methods reproduce $\rho(\lambda)$ rather well in the
peak regions. However, when we investigate in more detail the results for the ``valley'' region around $\lambda\!=\!0$, clear differences are
observed between the coarse-graining methods. Inset  B of Figure \ref{fig:Eq_cg} shows on a logarithmic scale the results for $\rho(\lambda)$ in this region, generated using the FFS method. All the coarse-graining methods deviate from the results of the full reaction network (ORN). The apparent effect of removing dimerisation (ED1/ED2) is to shift the minimum up in probability, with the macroscopic rate equation approach (ED1) having a stronger effect. Removing operator state fluctuations (EO) has the opposite effect, shifting the minimum down in probability. Methods EO-ED1 and EO-ED2 appear to show a combination of these two effects. Although these deviations from the ORN results are small, they will turn out to be rather important for the dynamical switching behaviour to be discussed in the next section. However,  if one is only interested in the steady state distribution, the choice of the particular coarse-graining method does not appear to be crucial.

\subsection{\large{Rate of stochastic switch flipping}}
\label{Sec:rate}

In many cases, fluctuation-driven dynamical properties are an
important output of a simulation of a biochemical network. This is
especially true of genetic switches, where a key characteristic is the
rate of flipping between stable states (as shown in Figure
\ref{fig:tracks_cg} for the model genetic switch).  When simulating these
systems, one requires not only an accurate representation of the steady-state distribution, but also of the dynamical behaviour of the system. We now test whether the various coarse-graining methods are able to reproduce the correct rate of stochastic flipping of the model switch. This is a particularly stringent test, since this fluctuation-driven process is likely to be highly sensitive to the accuracy with which dynamical fluctuations are reproduced in the different schemes. 

To measure the rate  $k_{\rm AB}$ of stochastic switch flipping, we use the FFS method \cite{Allen05,Allen06_1}, which allows the calculation of rate constants and the sampling of transition paths for rare events in stochastic dynamical systems. Because rare events (such as switch flipping) happen infrequently, standard simulations have difficulty in achieving good statistical sampling. In FFS, simulation  pathways
corresponding to the rare events are generated efficiently via a series
of interfaces, defined by a parameter $\lambda$, separating the initial and final states. At the same time, the rate constant is calculated
from the probabilities of reaching adjacent interfaces. FFS is described in more detail in Appendix \ref{app:FFS}, and in refs
\cite{Allen05,Allen06_1}. The method can also be used to obtain the steady state probability distribution as a function of the $\lambda$
parameter, as in inset B of Figure \ref{fig:Eq_cg} \cite{Valeriani07}. 

\begin{figure}[t]
\begin{center}
\includegraphics[width=0.5\columnwidth,clip=true]{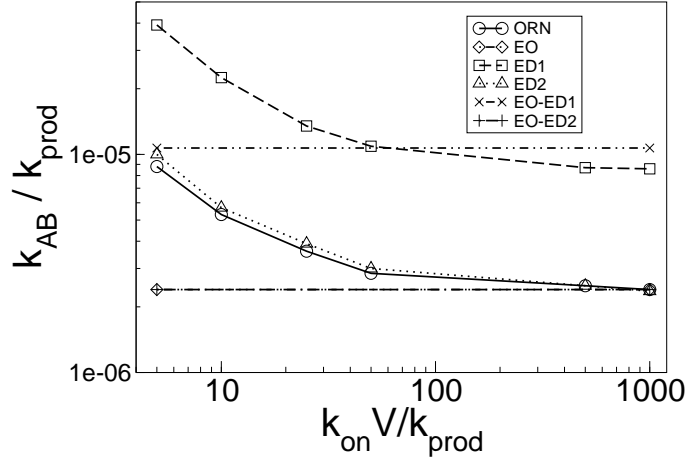}
\caption{Switch flipping rate $k_{\rm AB}$ as a function of the dimer-DNA association rate $k_{\rm on}$, adjusting $k_{\rm off}$ so that the
equilibrium constant for DNA binding remains unchanged. The switch is more stable when operator binding/unbinding is rapid, suggesting that
fluctuations in these reactions play an important role in the switch flipping. Methods that remove these reactions give a straight line in
the figure; among those, only methods EO and EO-ED2 are able to capture the asymptotic behavior of the curve for $k_{\rm on}\to\infty$. Method
ED2 always gives a good approximation of the rate, while ED1 consistently overestimates it by about one order of magnitude.
\label{fig:sw_rate1} }
\end{center}
\end{figure}

Figure \ref{fig:sw_rate1} shows the switch flipping rate $k_{\rm AB}$, as a function of the dimer-DNA association rate $k_{\rm on}$. The
dimer-DNA dissociation rate is adjusted to keep $K_{\rm D}^d\!=\!1$. The other parameters are fixed at $k_{\rm f}\!=\!5k_{\rm prod}$, $\mu\!=\!0.3
k_{\rm prod}$ and $K_{\rm D}^b\!=\!1/5$. For the full reaction network (ORN; solid line), the flipping rate decreases as DNA binding becomes
faster, reaching a plateau for very fast ($k_{\rm on}>500k_{\rm prod}$) operator association/dissociation. The switch is more stable when
operator binding/unbinding is rapid, suggesting that fluctuations in these reactions play an important role in switch flipping. For the EO
method, in which protein-DNA association/dissociation reactions are ``lost'', the flipping rate does not depend on $k_{\rm on}$ (since only
the equilibrium constant $K_{\rm D}^b$ features in this method and this is kept constant). As expected, the flipping rate for the EO method
corresponds to the ORN result in the limit of large $k_{\rm on}$. When we coarse-grain over protein-protein interactions (ED1 and ED2), our
results are very dependent on whether the macroscopic rate equations or the master equation is used to compute propensity functions. For the
rate equation approach (ED1), the decrease in $k_{\rm AB}$ with $k_{\rm on}$ is reproduced, but the switch is almost an order of magnitude
less stable than for the full reaction set. However, when the chemical master equation is used to compute the propensities, the results are
remarkably accurate - the ED2 approach gives switch flipping rates in good agreement with the full reaction set. The two methods EO-ED1 and
EO-ED2, which coarse-grain over both DNA binding and dimerisation reactions, also show this behaviour: for EO-ED1, where the rate equation
approximation is used, the switch flipping rate is similarly almost an order of magnitude too high, whereas for  EO-ED2, where the master
equation is used, $k_{\rm AB}$ is indistinguishable from that given by method EO (where dimerisation is simulated explicitly). These results
show that dimer-DNA binding plays an important role in switch flipping for association/dissociation rates in the physiological range, and
that, for reliable coarse-graining, effective propensities need to be computed with the master equation rather than the macroscopic rate
equation approximation.

\begin{figure}[t]
\begin{center}
\includegraphics[width=0.5\columnwidth,clip=true]{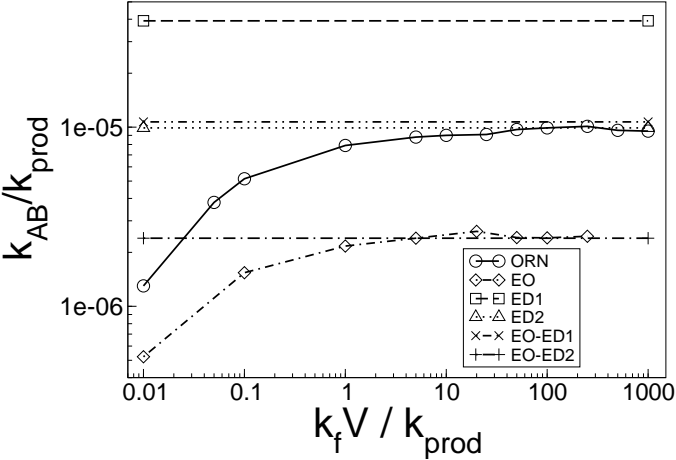}
\caption{Switch flipping rate $k_{\rm AB}$ as a function of the protein-protein association rate $k_{\rm f}$, adjusting $k_{\rm b}$ so that
the equilibrium constant for dimerisation remains unchanged. For the full reaction set, the switching rate increases with the rate of
dimerisation. The EO curve consistently underestimates the rate by approximately an order of magnitude. The methods that remove the
dimerisation reactions give a constant horizontal line. Among them, only methods EO-ED1 and EO give good results for high $k_{\rm f}$,
although for the latter we suspect this arises from a lucky cancellation of errors. 
 \label{fig:sw_rate2}}
\end{center}
\end{figure}

Figure \ref{fig:sw_rate2} shows the equivalent result when the
monomer-monomer association rate $k_{\rm f}$ is varied, adjusting
$k_{\rm b}$ so that $K_{\rm D}^d\!=\!1$. The other parameters are $k_{\rm
  on}\!=\!5k_{\rm prod}$, $\mu\!=\!0.3 k_{\rm prod}$ and $K_{\rm
  D}^b\!=\!1/5$. For the full reaction set (ORN), $k_{\rm AB}$ increases
with $k_{\rm f}$: as the dimerisation reactions become faster, the
switch becomes less stable. This is in contrast to the behaviour
observed in Figure \ref{fig:sw_rate1}. It appears that switch flipping
is hindered by fluctuations in the monomer-dimer reactions. This
apparently somewhat counter-intuitive result can perhaps be explained
as follows: protein is produced in the monomer form. To flip the switch, it needs to dimerise and bind to
the operator. If dimerisation is slow, the monomer may be degraded
before it has a chance to dimerise, and in this case it does not
contribute to flipping the switch. On the other hand, if dimerisation
is fast, then every monomer that is produced makes a contribution to
the dimer pool and can potentially bind to the operator, leading to
switch flipping. The EO approach (eliminating dimer-DNA binding) shows
the same increase in $k_{\rm AB}$ with $k_{\rm f}$, but underestimates
the value of $k_{\rm AB}$ by about an order of magnitude. This
supports our view that fluctuations in operator binding are important
for switch flipping. On eliminating dimerisation fluctuations (ED1 and
ED2), we observe the same problem with the macroscopic rate equation
approximation as in Figure \ref{fig:sw_rate1} - ED1 produces a
flipping rate that is too high, while ED2, where the master equation
is used, gives good agreement with the full reaction set (ORN) in the
high $k_{\rm f}$ limit. When both protein-protein and protein-DNA
association/dissociation reactions are eliminated, method EO-ED2 gives
results in agreement with EO in the high $k_{\rm f}$ limit. Method
EO-ED1 gives unexpectedly good results, in fairly close agreement with
ED2 and ORN. However, given that we expect removing DNA binding to
reduce the rate constant, while using the macroscopic rate equation
approximation increases it, this is likely to be just a lucky
cancellaton of errors for this particular parameter set. Figure
\ref{fig:sw_rate2} therefore demonstrates that fluctuations in the
monomer-monomer association/dissocation reactions actually disfavour
switch flipping. Moreover, as for Figure \ref{fig:sw_rate1}, we see that
the macroscopic rate equation approximation is not reliable for
predicting switch flipping rates, while coarse-graining over the
dimerisation reactions using the master equation approach (ED2) becomes
reliable when $k_{\rm f} > 5 k_{\rm{prod}}$ (for this parameter set).

We have also tested the various coarse-graining approaches for the
case where both protein-protein and protein-DNA
association/dissociation reactions are fast ($k_{\rm f}\!=\!100k_{\rm
  prod},k_{\rm on}\!=\!100k_{\rm prod}$). In this case, as expected,
methods EO, ED2 and EO-ED2 all give flipping rates in agreement with
the full reaction set (ORN), while ED1 and EO-ED1 do not. This
indicates once again that in general the macroscopic rate equation approach is
not suitable for computing switching rates.

\subsection{\large{Computational Performance}}
We measure the computational speed-up gained by the various
coarse-graining approximations. Of course, this depends on the
values that we choose for the rate constants. Table
\ref{tab:perf_eq} shows the CPU time, in seconds (on an AMD Athlon
1600+ processor), required for a simulation run of length $10^5k_{\rm prod}^{-1}$, for various values of the rate
constants $k_{\rm on}$ for protein-DNA binding and $k_{\rm f}$ for
dimerisation. In all cases, $k_{\rm off}$ and $k_{\rm b}$ are
adjusted so as to keep the equilibrium constants $K_{\rm D}^d$ and
$K_{\rm D}^b$ fixed. Considering the full reaction network (ORN) - top
row in the table - we observe that the CPU time is much more
sensitive to the dimerisation rate than to the protein-DNA binding
rate. This confirms that the SSA mostly executes monomer-monomer
association and dimer dissociation reactions, even when $k_{\rm on}$
is much greater than $k_{\rm f}$. This is because the
propensity for dimerisation depends on (roughly) the square of the
number of free monomers, which is generally quite
large. Protein-protein association/dissociation is therefore the
performance bottleneck for this system. Bearing this in mind, it is
not surprising that when we consider the next row in Table
\ref{tab:perf_eq}, we see that removing protein-DNA association and
dissociation reactions (EO), is only useful when the rate constants
for these reactions are exceedingly large.  Eliminating the protein-protein
association and dissociation reactions (ED1 and ED2) results in a
dramatic speed-up compared to the ORN case (rows 3 and 4). 
This speed-up is most impressive when the dimerisation rate is high. There is no
significant difference in CPU time between the ED1 and ED2
methods, as solving the dimerisation master equation can be done analytically and takes only a negligible time. 
When we eliminate both protein-DNA and protein-protein
association and dissociation (bottom two rows in Table
\ref{tab:perf_eq}), we obtain a further factor 2-25 increase in
speed. Again, there is no significant difference in CPU time between
methods EO-ED1 and EO-ED2. In this case, the solution of the master equation in method EO-ED2 is done 
numerically, and it can take up to a few hours. However this procedure is performed in a separate simulation
and the results are stored in a lookup table, to be used for all the simulations of the switch. Therefore,
we do not include the time needed to generating this table in Table \ref{tab:perf_eq}
We conclude that, in the
physiological parameter range, some computational speed-up can be
obtained by removing protein-DNA binding reactions; however, much more
computer time can be saved by coarse-graining protein-protein association
and dissociation reactions.


\begin{table}[b]
\begin{center}
\begin{tabular}{|c|c|c|c|c|}
\hline
 & $k_{\rm f}=5$    & $k_{\rm f}=100$ & $k_{\rm f}=5$        & $k_{\rm f}=100$  \\
 & $k_{\rm on}=5$ & $k_{\rm on}=5$  & $k_{\rm on}=100$ & $k_{\rm on}=100$ \\
\hline
ORN     & 5.81  & 113   & 5.55 & 118    \\
EO     & 5.25  & 103   & 5.08 & 70.7   \\
ED1    & 0.18  & 0.19  & 1.94 & 1.94   \\
ED2    & 0.18  & 0.19  & 1.92 & 1.91   \\
EO-ED1 & 0.085 & 0.082 & 0.081 & 0.081 \\
EO-ED2 & 0.083 & 0.082 & 0.081 & 0.084 \\
\hline
\end{tabular}
\end{center}
\caption{ CPU time (in seconds) required to simulate the system for
$t_{\rm sim}\!=\!10^5k_{\rm prod}^{-1}$, for different parameter sets.
Simulations were performed on an AMD Athlon 1600+ processor. The
dissociation rates were scaled such that the equilibrium constants
for dimerisation and operator binding were kept constant at $K_{\rm
D}^d\!=\!1/5$ and $K_{\rm D}^b\!=\!1$.} \label{tab:perf_eq}
\end{table}


\section{\large{Discussion}}

Understanding cellular control systems will require the study of very complex biochemical reaction networks. Computer simulations clearly have an important contribution to make in this area, since they can provide quantitative
understanding of how biochemical networks work. It is clear that
in many cases (including gene regulation), stochastic simulations
are required. However, the more complex the biochemical network is,
the more computationally expensive it is to simulate. Eliminating fast
reactions will be essential for simulating biochemical networks of the
scale and complexity that is relevant for biological cells. It is
therefore very important to understand how this can be done reliably, while
preserving the correct dynamical features of the full reaction
network. In this paper, we have made a systematic study of the
computational speedup and accuracy of a range of coarse-graining
schemes, for a model gene regulatory network. All gene regulatory
networks involve protein-protein and protein-DNA interactions. These
tend to be rapid in comparison to protein production (transcription,
translation and folding) and removal from the cell (active degradation
and dilution due to growth and division). We try to address the
general question of what the consequences are of eliminating
protein-protein or protein-DNA association and dissociation reactions
from stochastic simulations of gene regulatory networks. We use as our
case study a bistable genetic switch, since this gives us a very
sensitive readout, in the form of the switch flipping rate, of the
accuracy with which dynamical fluctuations are reproduced by the
various coarse-graining schemes. We hope that our results will prove
relevant to simulations of real genetic switches and gene regulatory
networks in general.

To coarse-grain the reaction scheme for the model genetic switch, within the context of Gillespie's Stochastic Simulation Algorithm (SSA), we
have used the approach described by Bundschuh {\em{et al.}} \cite{Bundschuh03}. Here, the reaction set is divided into ``fast'' and ``slow'' reactions. Chemical species whose number is changed by the fast reactions are designated ``fast''. A set of ``slow'' chemical species is constructed, which consists of original species that were unaffected by the fast reactions, together with new species, formed from linear combinations of the fast species, such that their number is unaffected by the fast reactions. The slow reactions are then rewritten in terms of the set of slow species, with effective propensity functions that depend on averages (and in some cases variances) of the fast reaction set, for fixed numbers of molecules of the slow species. These averages may be obtained by explicit or numerical solution of the chemical master equation for the fast reactions. Alternatively, one may make the approximation that the averages are well represented by the steady-state solutions of the corresponding macroscopic rate equations for the fast reactions. In either case, having computed the effective propensity functions, one simply implements the SSA for the slow reaction set, propagating the set of slow variables, using these effective propensities. 

For the model genetic switch, we investigated the effects of
eliminating protein-protein association/dissociation reactions, and/or
protein-DNA association/dissociation reactions, from the full reaction
set. We also compared the macroscopic rate equation approximation to the master equation approach for computing the effective
propensities. Using all the coarse-graining schemes, we computed the
steady-state probability distribution as well as spontaneous switch
flipping rates. We found that all the coarse-graining methods gave
good agreement with the full reaction network for the steady-state
probability distribution, although small deviations were observed
around the unstable steady state. However,
dramatic differences were observed in the switch flipping rates
computed using the different coarse-graining schemes. Elimination of
protein-DNA association/dissociation increased the stability of the
switch (but agreed with the full reaction set in the fast reaction
limit). In contrast, elimination of protein-protein association/dissociation decreased the stability of the switch (again, agreeing with the full reaction set in the fast reaction
limit). However, over most of the range of parameters tested, protein-protein
association/dissociation reactions can be eliminated with minimal effect on switching rates, and with the advantage of an impressive computational speed-up. This result is likely to prove very useful when simulating complex and computationally expensive networks. The implications of these observations for the physics of the switching mechanism for this model switch will be investigated in
a future publication \cite{morelli_prep}.

We also observed that the macroscopic rate equation approximation does not produce reliable switching rates, even though the steady-state probability distribution is reasonably well reproduced. Typically, switching rates computed using the macroscopic rate equation approximation are an order of magnitude too high, even in the limit of fast reactions. In contrast, when the chemical master equation is used to compute the effective propensities, results are in excellent agreement with the full reaction set for fast reaction rates. These results serve as a warning that care must be taken in how coarse-graining is applied. As an example, the lysogeny-lysis switch of bacteriophage lambda is extremely stable to fluctuations \cite{Little99,Aurell02}, a fact that computational modelling (using macroscopic approximations) has  thus far been unable satisfactorily to explain
\cite{AS,Aurell02}. In such a case, careful
coarse-graining is crucially important.  

Our results show that under certain biologically relevant conditions fast reactions can be
eliminated while preserving the correct dynamical characteristics of
the system, even when highly sensitive fluctuation-driven quantities
such as switch flipping rates are considered. This is very encouraging for the simulation of more complex reaction networks, and it would be interesting to apply these approaches to   more
complicated genetic switches, and also for other gene regulatory
networks where dynamical fluctuations are important. We hope that this
work will be of use as a ``tutorial'' in designing and implementing
coarse-graining schemes, and that it may aid in pointing the way to
accurate and efficient coarse-grained simulations of a wide variety of
interesting and important biochemical networks.

\begin{acknowledgments}
The authors are grateful to Patrick Warren for pointing out reference 5 and for very valuable discussions. This work is part of the research program of the
"Stichting voor Fundamenteel Onderzoek der Materie (FOM)", which is financially supported by the "Nederlandse organisatie voor
Wetenschappelijk Onderzoek (NWO)''. R.J.A. was funded by the European Union Marie Curie program and by the Royal Society of Edinburgh.
\end{acknowledgments}

\newpage

\appendix

\section{\large{Solving the operator binding Master Equation}} \label{app:ME2}
In the EO approach, instead of solving the macroscopic rate equation for operator binding, one can solve the corresponding chemical Master
Equation. However, as the operator states can be present only in copy number 0 or 1, the state space is extremely limited, and the solution
of the Master Equation coincides with the solution for the rate equation - as we demonstrate here.

The Master Equation for reactions (b) in Table  \ref{tab:reactions} is the following:
\begin{eqnarray}\label{eq:ME2}
\frac{\partial}{\partial t}\, p(n_{A_2},n_{B_2})  & = & \\ \nonumber 
& - & p(n_{A_2},n_{B_2})(k_{\rm on}n_On_{\rm A_2} + k_{\rm on}n_On_{\rm B_2}) \\ \nonumber  
& - & p(n_{A_2},n_{B_2})( k_{\rm off}n_{O{\rm A_2}} + k_{\rm off}n_{O{\rm B_2}}\\ \nonumber 
& + & p(n_{A_2}-1,n_{B_2})k_{\rm off}(n_{O{\rm A_2}}+1)  \\ \nonumber 
& + & p(n_{A_2},n_{B_2}-1)k_{\rm off}(n_{O{\rm B_2}}+1)  \\ \nonumber
& + & p(n_{A_2}+1,n_{B_2})k_{\rm on}(n_{O}+1)(n_{\rm A_2}-1) \\ \nonumber
& + & p(n_{A_2},n_{B_2}+1)k_{\rm on}(n_{O}+1)(n_{\rm B_2}-1). 
\end{eqnarray}
Only three states are possible: $(O\!=\!1,{\rm A_2},{\rm B_2})$, $(O{\rm A_2}\!=\!1,{\rm A_2}\!-\!1,{\rm B_2})$ and 
$(O{\rm B_2}\!=\!1,{\rm A_2},{\rm B_2}\!-\!1)$. This greatly simplifies Eq. ({\ref{eq:ME2}):
\begin{eqnarray}\label{eq:ME2_2}
p(n_{A_2},n_{B_2})k_{\rm on}(n_{\rm A_2}\!+\!n_{\rm A_2}) & = & \\ \nonumber
p(n_{A_2}\!-\!1,n_{B_2}) k_{\rm off} \!+\! p(n_{A_2},n_{B_2}\!-\!1)k_{\rm off}, \\
\nonumber p(n_{A_2},n_{B_2}\!-\!1) & = & p(n_{A_2},n_{B_2})k_{\rm on} n_{\rm B_2}, \\ \nonumber
p(n_{A_2}\!-\!1,n_{B_2}) & = & p(n_{A_2},n_{B_2})k_{\rm on} n_{\rm A_2}. 
\end{eqnarray}
The solutions of Eq. (\ref{eq:ME2_2}) can be easily computed:
\begin{eqnarray}
\langle n_{O}\rangle_{\mathrm{\hat{A}_2},\mathrm{\hat{B}_2}}^{\mathrm{ME}} & = & p(n_{A_2},n_{B_2})  =  \frac {1}{1+(K_D^b)^{-1}(n_{\rm A_2}+n_{\rm B_2})}, \\ \nonumber
\langle n_{O\rm A_2}\rangle_{\mathrm{\hat{A}_2},\mathrm{\hat{B}_2}}^{\mathrm{ME}} & = & p(n_{A_2}\!-\!1,n_{B_2}) =  \frac {(K_D^b)^{-1}n_{\rm A_2}}{1+(K_D^b)^{-1}(n_{\rm A_2}+n_{\rm B_2})}, \\ \nonumber
\langle n_{O\rm B_2}\rangle_{\mathrm{\hat{A}_2},\mathrm{\hat{B}_2}}^{\mathrm{RE}} & = & p(n_{A_2},n_{B_2}\!-\!1) =  \frac {(K_D^b)^{-1}n_{\rm B_2}}{1+(K_D^b)^{-1}(n_{\rm A_2}+n_{\rm B_2})}. 
\end{eqnarray}

\section{\large{Solving the dimerisation Master Equation}} \label{app:ME1}
Following the approach described in \cite{Bundschuh03}, the Master Equation for the reaction $X+X\rightleftharpoons X_2$,
with rate constants $k_{\rm f}$ for association and $k_{\rm b}$ for dissociation, with system volume $V$, and given a total number or
monomers+dimers $n_{X_T}$, where $n_{X_T}=n_{X}+2n_{X_2}$, is the following:

\begin{eqnarray}\label{eq:me}
&& \frac{\partial}{\partial t}\, p(n_{X_2}|n_{X_T}) = k_{\rm b}(n_{X_2}+1)\: p(n_{X_2}+1|n_{X_T})  -  \\ \nonumber
&& \left[\frac{k_{\rm f}(n_{X_T}-2n_{X_2})(n_{X_T}-2n_{X_2}-1)}{2V}+k_{\rm b}n_{X_2} \right]\:p(n_{X_2}|n_{X_T}) +  \\ \nonumber &&
\frac{k_{\rm f}(n_{X_T}-2n_{X_2}+2)(n_{X_T}-2n_{X_2}+1)}{2V}\: p(n_{X_2}-1|n_{X_T})
\end{eqnarray}

Eq. (\ref{eq:me}) can be solved numerically in steady state (starting from an initial guess $n_{X_2}=0$), to obtain the exact probability distribution for
the number of
dimers $n_{X_2}$, for a given total number $n_{X_T}$ of momoners+dimers. The probability distribution for the monomer number can be trivially
obtained from the dimer distribution noting that $p(n_{X}|n_{X_T})=n_{X_T}-2p(n_{X_2}|n_{X_T})$. Eq. (\ref{eq:me}) is solved for a range of
values of $n_{X_T}$; results are stored in look-up tables, which are later used to compute effective propensities for the coarse-grained
simulations.

\section{\large{Forward Flux Sampling}} \label{app:FFS}
Forward Flux Sampling (FFS) \cite{Allen05,Allen06_1,Allen06_2} is a method for sampling  spontaneous transitions between two regions in phase space
$A$ and $B$, and for computing the rate constant for such transitions. A and B are defined by an order parameter $\lambda$, such that $\lambda<\lambda_{\rm A}$ in A and $\lambda>\lambda_{\rm B}$ in B. A series of nonintersecting
surfaces $\lambda_0,\cdots ,\lambda_n$ are defined in phase space, such that $\lambda_0 = \lambda_{\rm A}$ and $\lambda_n=\lambda_{\rm B}$, and such that any path from A to B must cross each interface, without reaching $\lambda_{i+1}$ before $\lambda_i$. The transition rate $k_{\rm AB}$ from A to B is the average flux $\bar{\Phi}_{\rm{A},n}$ of trajectories reaching B from A. This can be decomposed in the following way:
\begin{equation}\label{eq:ffs1}
k_{\rm AB}=\bar{\Phi}_{\rm{A},n}=\bar{\Phi}_{\rm{A},0}P(\lambda_n|\lambda_0)=\bar{\Phi}_{{\rm A},0}\prod_{i=0}^{n-1}P(\lambda_{i+1}|\lambda_i).
\end{equation}
Here, $\bar{\Phi}_{{\rm A},0}$ is the average flux of trajectories leaving A in the direction of B and $P(\lambda_n|\lambda_0)$ is the
probability that a trajectory that crosses $\lambda_0$ in the direction of B will eventually reach B before returning to A. On the right-hand
side, $P(\lambda_{i+1}|\lambda_i)$ is the probability that a trajectory which reaches  $\lambda_i$, having come from A, will reach
$\lambda_{i+1}$ before returning to A. In Eq.(\ref{eq:ffs1}), the flux of trajectories from A to B is split into the flux across the first
interface $\lambda_0$, multiplied by the probability of getting from that
interface to B, without returning to A. This last term is then factorized in a product of conditional probabilities of reaching the next interface (before returning to A), having arrived at a particular interface from A. We note that this does not imply a Markov approximation \cite{Allen06_1}.

The flux  $\bar{\Phi}_{{\rm A},0}$ is obtained by running a simulation of the system in the ``basin of attraction'' of A and counting how many times the trajectory in phase space crosses
$\lambda_0$ coming from A. At the same time, one generates a collection of phase space points that correspond to the moments that the trajectory reached $\lambda_0$, moving in the direction of B. This collection of points is then used as the starting point for a calculation of $P(\lambda_{1}|\lambda_0)$. A point from the collection is chosen at random and used to initiate a new trajectory, which is continued until either $A$ or $\lambda_1$ is reached. If $\lambda_1$ is reached, the trial is designated a ``success''. This is repeated many times, generating an estimate for $P(\lambda_{1}|\lambda_0)$ (the number of successes divided by the total number of trials), plus a new collection of points at $\lambda_1$ that are the end points of the successful trial runs. This collection of points is used to initiate trial runs to $\lambda_2$, generating an estimate for $P(\lambda_{2}|\lambda_1)$ and a new collection at $\lambda_2$, etc. FFS also allows sampling of the trajectories corresponding to the transition (the transition path ensemble) by tracing back to A paths that eventually arrive in B. Further details on FFS are given in \cite{Allen05,Allen06_1}. The use of FFS to compute stationary probability distributions is demonstrated in \cite{Valeriani07}.

\newpage
\renewcommand{\arraystretch}{0.5}
\normalsize

\begin{tabular}{cccccc}
&$\quad\qquad \mathrm{Reaction} \quad\qquad$ & $\qquad\quad\mathrm{Propensity}\qquad\quad$ & $\qquad\quad\mathrm{Reaction}\quad\qquad$  & $\quad\qquad\mathrm{Propensity}\quad\qquad$ & \\
\hline
{\rm Dimerisation} &$\mathrm{A}+\mathrm{A} \rightleftharpoons \mathrm{A}_2$  & $k_{\rm f}n_{\mathrm{A}}(n_{\mathrm{A}}\!-\!1),\quad k_{\rm b}n_{\mathrm{A}_2}$
&$\mathrm{B}+\mathrm{B} \rightleftharpoons \mathrm{B}_2$  & $k_{\rm f}n_{\mathrm{B}}(n_{\mathrm{B}}\!-\!1),\quad k_{\rm b}n_{\mathrm{B}_2}$ &a\\
{\rm DNA\:binding} &${O} + \mathrm{A}_2 \rightleftharpoons {O}\mathrm{A}_2$  & $k_{\rm on} n_{{O}}n_{\mathrm{A}_2},\qquad k_{\rm
off}n_{O\mathrm{A}_2}$  & ${O} + \mathrm{B}_2 \rightleftharpoons O\mathrm{B}_2$  & $k_{\rm on}
n_{{O}}n_{\mathrm{B}_2},\qquad k_{\rm off}n_{O\mathrm{B}_2}$ &b\\
{\rm Production} &${O} \to {O} + \mathrm{A}$ & $k_{\rm prod}n_{{O}}$ & ${O} \to {O} + \mathrm{B}$ & $k_{\rm prod}n_{{O}}$ &c\\
{\rm Production} &${O}\mathrm{A}_2 \to {O}\mathrm{A}_2 + \mathrm{A}$  & $k_{\rm prod}n_{O\mathrm{A}_2}$  &
${O}\mathrm{B}_2 \to {O}\mathrm{B}_2 + \mathrm{B}$ & $k_{\rm prod}n_{O\mathrm{B}_2}$  &d\\
{\rm Degradation} &$\mathrm{A} \to \emptyset$ & $\mu n_{\mathrm{A}}$  & $\mathrm{B} \to \emptyset$  & $\mu n_{\mathrm{B}}$ &e\\
\end{tabular}

\vspace{1cm}

Table 1: Reactions and propensity functions for the model genetic switch.


\newpage

\begin{center}
\begin{tabular}{|c|c|c|c|c|c|}
\hline
Name & Reaction & Propensity & Method & Coarse-grained & Definition\\
 &  &  &  & variable & \\

\hline\hline
EO & $\emptyset \to \mathrm{A}$ & $k_{\rm prod}\langle n_{O}+n_{O\rm A_2}\rangle_{\mathrm{\hat{A}_2},\mathrm{\hat{B}_2}}^{\mathrm{RE}}$ & Rate Equation &
$\mathrm{\hat{A}_2}$ & $n_{\mathrm{\hat{A}_2}} = n_{\mathrm{A}_2}+n_{O\mathrm{A}_2}$ \\
  & $\emptyset \to \mathrm{B}$ &  $k_{\rm prod}\langle n_{O}+n_{O\rm B_2}\rangle_{\mathrm{\hat{A}_2},\mathrm{\hat{B}_2}}^{\mathrm{RE}}$ & & $\mathrm{\hat{B}_2}$ & $n_{\mathrm{\hat{B}_2}} =
  n_{\mathrm{B}_2}+n_{O\mathrm{B}_2}$ \\
  & $\mathrm{A}+\mathrm{A}\rightleftharpoons \mathrm{\hat{A}_2}$ & $k_{\rm on} n_{\rm A}(n_{\rm A}\!-\!1),\; k_{\rm off}\langle n_{\rm
  A_2}\rangle_{\mathrm{\hat{A}_2},\mathrm{\hat{B}_2}}^{\mathrm{RE}}$ & &  &\\
  & $\mathrm{B}+\mathrm{B}\rightleftharpoons \mathrm{\hat{B}_2}$ & $k_{\rm on} n_{\rm B}(n_{\rm B}\!-\!1),\; k_{\rm off}\langle n_{\rm
  B_2}\rangle_{\mathrm{\hat{A}_2},\mathrm{\hat{B}_2}}^{\mathrm{RE}}$ & &  &\\
  & $\mathrm{A} \to \emptyset$ & $\mu n_{\rm A}$ & &  &\\
  & $\mathrm{B} \to \emptyset$ & $\mu n_{\rm B}$ & &  &\\
\hline
ED1 & ${O}+2\check{\mathrm{A}}\rightleftharpoons O\mathrm{A}_2$ & $k_{\rm on}\langle n_{\rm
A_2}\rangle_{\check{\mathrm{A}}}^{\mathrm{RE}},\; k_{\rm off} n_{O\rm
  A_2}$ & Rate Equation & $\check{\mathrm{A}}$ & $n_{\check{\mathrm{A}}} = n_{\mathrm{A}}+2n_{\mathrm{A}_2}$  \\
  & ${O}+2\check{\mathrm{B}}\rightleftharpoons O\mathrm{B}_2$ & $k_{\rm on}\langle n_{\rm B_2}\rangle_{\check{\mathrm{B}}}^{\mathrm{RE}}, \; k_{\rm off} n_{O\rm
  B_2}$ & &  $\check{\mathrm{B}}$ & $n_{\check{\mathrm{B}}} =n_{\mathrm{B}}+2n_{\mathrm{B}_2}$ \\
  & ${O} \to {O} +\check{\mathrm{A}}$ & $k_{\rm prod} n_{O}$   & &  &\\
  & ${O} \to {O} +\check{\mathrm{B}}$ & $k_{\rm prod} n_{O}$  & &  &\\
  & $O\mathrm{A}_2 \to O\mathrm{A}_2 + \check{\mathrm{A}}$ & $k_{\rm prod} n_{O\rm A_2}$ & &  &\\
  & $O\mathrm{B}_2 \to O\mathrm{B}_2 + \check{\mathrm{B}}$ & $k_{\rm prod} n_{O\rm B_2}$ & &  &\\
  & $\check{\mathrm{A}} \to \emptyset$ & $\mu \langle n_{\rm A}\rangle_{\check{\mathrm{A}}}^{\mathrm{RE}}$ & &  &\\
  & $\check{\mathrm{B}} \to \emptyset$ & $\mu \langle n_{\rm B}\rangle_{\check{\mathrm{B}}}^{\mathrm{RE}}$ & &  &\\
\hline
ED2 & ${O}+2\check{\mathrm{A}}\rightleftharpoons O\mathrm{A}_2$ & $k_{\rm on}\langle n_{\rm
A_2}\rangle_{\check{\mathrm{A}}}^{\mathrm{ME}}, \;k_{\rm off}
n_{O\rm A_2}$ & Master Equation & $\check{\mathrm{A}}$ &$n_{\check{\mathrm{A}}} =n_{\mathrm{A}}+2n_{\mathrm{A}_2}$ \\
  & ${O}+2\check{\mathrm{B}}\rightleftharpoons O\mathrm{B}_2$ & $k_{\rm on}\langle n_{\rm B_2}\rangle_{\check{\mathrm{B}}}^{\mathrm{ME}}, \; k_{\rm off} n_{O\rm
  B_2}$ & & $\check{\mathrm{B}}$ &$n_{\check{\mathrm{B}}} =n_{\mathrm{B}}+2n_{\mathrm{B}_2}$\\
  & ${O} \to {O} +\check{\mathrm{A}}$ & $k_{\rm prod} n_{O}$ & &  &\\
  & ${O} \to {O} +\check{\mathrm{B}}$ & $k_{\rm prod} n_{O}$ & &  &\\
  & $O\mathrm{A}_2 \to O\mathrm{A}_2 + \check{\mathrm{A}}$ & $k_{\rm prod} n_{O\rm A_2}$ & &  &\\
  & $O\mathrm{B}_2 \to O\mathrm{B}_2 + \check{\mathrm{B}}$ & $k_{\rm prod} n_{O\rm B_2}$ & &  &\\
  & $\check{\mathrm{A}} \to \emptyset$ & $\mu \langle n_{\rm A}\rangle_{\check{\mathrm{A}}}^{\mathrm{ME}}$ & &  &\\
  & $\check{\mathrm{B}} \to \emptyset$ & $\mu \langle n_{\rm B}\rangle_{\check{\mathrm{B}}}^{\mathrm{ME}}$ & &  &\\
\hline
EO-ED1 & $\emptyset \to \tilde{\mathrm{A}}$ & $k_{\rm prod}\langle n_{O}+n_{O\rm
A_2}\rangle_{\tilde{\mathrm{A}},\tilde{\mathrm{B}}}^{\mathrm{RE}}$ & Rate Equation &$\tilde{\mathrm{A}}$ &
$n_{\tilde{\mathrm{A}}} = n_{\mathrm{A}}+2n_{\mathrm{A}_2}+2n_{O\mathrm{A}_2}$\\
  & $\emptyset \to \tilde{\mathrm{B}}$ & $k_{\rm prod}\langle n_{O}+n_{O\rm
  B_2}\rangle_{\tilde{\mathrm{A}},\tilde{\mathrm{B}}}^{\mathrm{RE}}$ & & $\tilde{\mathrm{B}}$ &
  $n_{\tilde{\mathrm{B}}} = n_{\mathrm{B}} +2n_{\mathrm{B}_2}+2n_{O\mathrm{B}_2}$\\
  & $\tilde{\mathrm{A}} \to \emptyset$ & $\mu \langle n_{\rm A}\rangle_{\tilde{\mathrm{A}},\tilde{\mathrm{B}}}^{\mathrm{RE}}$ & &  &\\
  & $\tilde{\mathrm{B}} \to \emptyset$ & $\mu \langle n_{\rm B}\rangle_{\tilde{\mathrm{A}},\tilde{\mathrm{B}}}^{\mathrm{RE}}$ & &  &\\
\hline
EO-ED2 & $\emptyset \to \tilde{\mathrm{A}}$ & $k_{\rm prod}\langle n_{O}+n_{O\rm
A_2}\rangle_{\tilde{\mathrm{A}},\tilde{\mathrm{B}}}^{\mathrm{ME}}$ & Master Equation &$\tilde{\mathrm{A}}$ &
$n_{\tilde{\mathrm{A}}} = n_{\mathrm{A}}+2n_{\mathrm{A}_2}+2n_{O\mathrm{A}_2}$ \\
  & $\emptyset \to \tilde{\mathrm{B}}$ & $k_{\rm prod}\langle  n_{O}+n_{O\rm
  B_2}\rangle_{\tilde{\mathrm{A}},\tilde{\mathrm{B}}}^{\mathrm{ME}}$ & & $\tilde{\mathrm{B}} $ &
  $n_{\tilde{\mathrm{B}}} = n_{\mathrm{B}} +2n_{\mathrm{B}_2}+2n_{O\mathrm{B}_2}$\\
  & $\tilde{\mathrm{A}} \to \emptyset$ & $\mu \langle n_{\rm A}\rangle_{\tilde{\mathrm{A}},\tilde{\mathrm{B}}}^{\mathrm{ME}} $ & &  &\\
  & $\tilde{\mathrm{B}} \to \emptyset$ & $\mu \langle n_{\rm B}\rangle_{\tilde{\mathrm{A}},\tilde{\mathrm{B}}}^{\mathrm{ME}} $ & &  &\\
\hline
\end{tabular}
\end{center}

\vspace{1cm}

{Table 2: Summary of coarse-graining schemes for the original reaction set
(Table \ref{tab:reactions}): eliminating operator binding (EO), eliminating dimerisation reactions using the Macroscopic Rate Equation (ED1) or the Master
Equation (ED2), eliminating both dimerisation and operator binding using the Macroscopic Rate Equation (EO-ED1) or the Master Equation
(EO-ED2). For each coarse-graining scheme, the coarse-grained reaction set is indicated together with the propensity function for each reaction. We also give definitions of the new slow variables for each scheme.
}

\newpage

\begin{tabular}{cccc}
$\quad\qquad\mathrm{Reaction}\qquad\quad$ &$\qquad\quad\mathrm{Propensity}\qquad\quad$ & $\qquad\quad\mathrm{Reaction}\qquad\quad$ & $\qquad\quad\mathrm{Propensity}\qquad\quad$\\
\hline
$\mathrm{A}+\mathrm{A} \rightleftharpoons \mathrm{A}_2$ & $k_{\rm f}n_{\mathrm{A}}(n_{\mathrm{A}}\!-\!1),\:k_{\rm b}n_{\mathrm{A}_2}$
& $\mathrm{B}+\mathrm{B} \rightleftharpoons \mathrm{B}_2$  & $k_{\rm f}n_{\mathrm{B}}(n_{\mathrm{B}}\!-\!1),\:k_{\rm
b}n_{\mathrm{B}_2}$ \\
${O} + \mathrm{A}_2 \rightleftharpoons {O}\mathrm{A}_2$ & $k_{\rm on} n_{{O}}n_{\mathrm{A}_2},\:k_{\rm
off}n_{O\mathrm{A}_2}$ & ${O} + \mathrm{B}_2 \rightleftharpoons {O}\mathrm{B}_2 $ & $k_{\rm on}
n_{{O}}n_{\mathrm{B}_2},\:k_{\rm off}n_{O\mathrm{B}_2}$
\end{tabular}

\vspace{1cm}

Table 3: Reaction scheme for the preliminary simulations to
  compute the effective propensity functions given in
  Eqs. (\ref{eq:a_eff}) and (\ref{eq:mu_eff}), for scheme EO-ED2.

\newpage

\begin{center}
\begin{tabular}{|c|c|c|c|c|}
\hline
 & $k_{\rm f}=5$    & $k_{\rm f}=100$ & $k_{\rm f}=5$        & $k_{\rm f}=100$  \\
 & $k_{\rm on}=5$ & $k_{\rm on}=5$  & $k_{\rm on}=100$ & $k_{\rm on}=100$ \\
\hline
ORN     & 5.81  & 113   & 5.55 & 118    \\
EO     & 5.25  & 103   & 5.08 & 70.7   \\
ED1    & 0.18  & 0.19  & 1.94 & 1.94   \\
ED2    & 0.18  & 0.19  & 1.92 & 1.91   \\
EO-ED1 & 0.085 & 0.082 & 0.081 & 0.081 \\
EO-ED2 & 0.083 & 0.082 & 0.081 & 0.084 \\
\hline
\end{tabular}
\end{center}

\vspace{1cm}

Table 4: CPU time (in seconds) required to simulate the system for
$t_{\rm sim}\!=\!10^5k_{\rm prod}^{-1}$, for different parameter sets.
Simulations were performed on an AMD Athlon 1600+ processor. The
dissociation rates were scaled such that the equilibrium constants
for dimerisation and operator binding were kept constant at $K_{\rm
D}^d\!=\!1/5$ and $K_{\rm D}^b\!=\!1$.

\newpage

\listoffigures

\newpage

\begin{center}
\includegraphics[width=\columnwidth]{501802JCP1.eps}
\vspace{1cm}
{M. J. Morelli et al, Figure 1}
\end{center}

\newpage

\begin{center}
\includegraphics[width=\columnwidth,clip=true]{501802JCP2.eps}
\vspace{1cm}
{M. J. Morelli et al, Figure 2}
\end{center}

\newpage

\begin{center}
\includegraphics[width=\columnwidth,clip=true]{501802JCP3.eps}
\vspace{1cm}
{M. J. Morelli et al, Figure 3}
\end{center}

\newpage

\begin{center}
\includegraphics[width=\columnwidth,clip=true]{501802JCP4.eps}
\vspace{1cm}
{M. J. Morelli et al, Figure 4}
\end{center}

\newpage

\begin{center}
\includegraphics[width=\columnwidth,clip=true]{501802JCP5.eps}
\vspace{1cm}
M. J. Morelli et al, Figure 5
\end{center}

\end{document}